\documentclass[aps,prd,a4paper,onecolumn,notitlepage,nofootinbib,groupedaddress,showpacs,showkeys]{revtex4-1}
\pdfoutput=1
\usepackage {amsmath}
\usepackage {amssymb}
\usepackage {amsfonts}
\usepackage {latexsym}
\usepackage {txfonts}
\usepackage {multirow}
\usepackage {hhline}
\usepackage {graphicx}
\usepackage {hyperref}
\usepackage {color}
\usepackage {bm}
\usepackage{appendix}
\usepackage{acronym}
\usepackage{dcolumn}   
\usepackage {url}
\usepackage{framed}
\usepackage{cleveref}
\usepackage{braket}

\newcommand{\dcc}{LIGO-P1300194-v2}

\newcommand\pvec{\vec{\theta}}
\newcommand\mvec{\vec{M}}
\newcommand{\Mc}{\mathcal{M}}
\newcommand{\Msun}{\ensuremath{\mathrm{M}_{\odot}}}
\newcommand{\A}{\mathcal{A}}
\newcommand{\coa}{{\phi_c}}
\newcommand{\ml}{{\rm ML}}
\newcommand{\hh}{\hat{h}}
\newcommand{\Dtheta}{\Delta\theta}
\newcommand{\ra}{\alpha}
\newcommand{\dec}{\delta}

\begin{document}

\title{Optimizing gravitational-wave searches for a population of coalescing binaries:\\Intrinsic parameters}
\author{T.~Dent}~\email{thomas.dent@aei.mpg.de}
\affiliation{Albert-Einstein-Institut (Max-Planck-Institut f{\"ur} Gravitationsphysik), Callinstr.~38, Hannover, Germany and Leibniz-Universit{\"a}t Hannover, Germany}
\author{J.~Veitch}~\email{john.veitch@ligo.org}
\affiliation{Nikhef, Science Park 105, Amsterdam 1098XG, Netherlands}
\date{\today, \dcc}

\begin{abstract}
\noindent
We revisit the problem of searching for gravitational waves from inspiralling compact binaries in 
Gaussian coloured noise. If the intrinsic parameters of a quasi-circular, non-precessing binary are known, 
then the optimal statistic for detecting the dominant mode signal in a single interferometer is given by the 
well-known two-phase matched filter.
However, the matched filter signal-to-noise ratio (SNR) is \emph{not} in general an optimal statistic for an 
astrophysical population of signals, since their distribution over the intrinsic parameters will almost 
certainly not mirror that of noise events, which is determined by the (Fisher) information metric.
Instead, the optimal statistic for a given astrophysical distribution will be the Bayes factor, which we 
approximate using the output of a standard template matched filter search. We then quantify the improvement 
in number of signals detected for various populations of non-spinning binaries: for a distribution of signals
uniformly distributed in volume and with component masses distributed uniformly over the range $1\leq m_{1,2}/
\Msun\leq 24$, $(m_1+m_2)/\Msun\leq 25$ at fixed expected SNR, we find $\gtrsim 20\%$ more signals at a false 
alarm threshold of $10^{-6}\,$Hz in a single detector. 
The method may easily be generalized to binaries with non-precessing spins.
\end{abstract}

\maketitle

\section{Introduction}
\noindent
Although some hints exist to the possible masses and spins of coalescing compact binary (CBC) systems visible 
to ground-based gravitational-wave (GW) interferometers \cite{LSCRates}, the dearth of current astronomical 
observations allows for a very broad uncertainty in the distribution of signals over these ``intrinsic'' 
parameters. Thus, searches for GWs from these sources must consider a correspondingly broad parameter space, 
which leads to increased computational cost and lower search efficiency relative to the case of binaries of 
known parameter values. 
Techniques for covering the space in a computationally efficient way are well developed (e.g.\ 
\cite{Babak:2006ty,Cokelaer:2007,Keppel:2013kia}), but the problem of dealing with the increased number of 
false alarms arising from a broad search space has received relatively little attention. Given the improved 
low-frequency sensitivity of Advanced LIGO and Virgo, and the probable need to search for spinning signals 
(see \cite{Brown:2012gs,Brown:2012qf,Ajith:2012mn,Harry:2013tca,Privitera:2013xza} and references therein) 
implying a yet larger parameter space, this issue is likely to become increasingly urgent. 

We build on methods used in previous searches for CBC signals, as described in detail in \cite{DuncanThesis,FindChirp,ihope,S5Year1}. 
In these, the intrinsic parameter space is searched by identifying a set (``bank'') of discrete parameter 
values, each representing a ``template'' signal, which are sufficiently closely spaced that the possible 
loss of search sensitivity due to a discrepancy between any true signal and the best fitting template is small 
and bounded above \cite{Balasubramanian:1995bm,Owen:1996,OwenSathya:1999}. The matched filter is evaluated for 
each template, and for each time sample in the experimental data \cite{DuncanThesis,FindChirp}.
Then the standard approach for detection of coalescing binaries with unknown intrinsic parameters is (at least in 
Gaussian noise) to weight all templates equally. This choice is implemented by taking the detection statistic, 
for candidate events in data from a single interferometric detector, to be the local maximum of a matched filter 
signal-to-noise ratio (SNR) over the bank, and over coalescence time. The SNR maximum corresponds to the 
maximum-likelihood value of the signal parameters in Gaussian noise~\cite{FinnChernoff:1993}. 
The resulting ``triggers'' from each detector are subsequently tested for consistency of parameter values between 
detectors, and the detection statistic for the resulting coincident events is taken to be the quadrature sum of 
single-detector SNRs (e.g.\ \cite{ihope}); however, we will not consider these multi-detector stages of the 
analysis here.

In this work we first argue (Section~\ref{sec:theoretical}) that the maximum matched filter SNR is \emph{not} 
in general an optimum detection statistic in the Neyman-Pearson sense of maximizing detection probability at 
fixed false alarm probability.  This lack of optimality is due to the necessity of marginalizing over the signal 
parameters when evaluating the Bayes factor (likelihood ratio) for the signal \emph{vs.}\ noise hypotheses. 
We show that an approximation of the required integral around a peak of signal 
likelihood yields, in addition to a strongly increasing dependence on the SNR, a non-trivial dependence on the 
intrinsic parameters. This dependence varies as the ratio of the prior probability density of signals to the 
density of the Fisher information metric for the filters (closely related to the metric used to lay out the 
template bank \cite{Balasubramanian:1995bm,Owen:1996,OwenSathya:1999}). 
Only when these two densities are equal is the matched filter SNR optimal. The use of the maximum SNR as a 
detection statistic over a broad parameter space thus corresponds to a prior over the binary's intrinsic 
parameters which is highly artificial, and may differ considerably from any astrophysically plausible 
distribution. For instance, the implied prior in a search for non-spinning binary inspirals with component masses 
1--24\,\Msun\ and maximum mass 25\,\Msun\ varies by several orders of magnitude over the space of component 
masses.\footnote{The density actually vanishes at the equal-mass line $m_1=m_2$, leading to an (integrable)
singularity in the implied prior, however even without considering templates near this line we see 4-5 orders of 
magnitude variation.}

Similarly, various detection statistics used in searches for unmodelled gravitational-wave bursts have been shown 
\cite{Searle:2008ap} to correspond to non-physical or implausible priors on the distribution of signals; and the 
maximization process involved in the ``F-statistic'' matched filter search for continuous gravitational waves 
corresponds to a non-physical prior on the source parameters for an individual template \cite{Prix:2009}.

In Section~\ref{sec:findchirp} we discuss implementation in the context of a matched filter search over a 
standard template bank; we are able to correct for the implicit prior with a negligible increase in 
computational cost, by computing a likelihood ratio statistic that effectively re-weights template matched 
filter values via a mass-dependent function incorporating both an assumed astrophysical prior and the Fisher 
metric density. 

Other approaches to using likelihood statistics for coalescing binary signals are described in 
\cite{Biswas:2012tv,Cannon:2012zt}. However, these studies are primarily concerned with accounting for the 
presence of non-Gaussian transient noise and/or variable detector sensitivity.
The algorithm described in \cite{Cannon:2012zt}, intended primarily to improve search performance in 
non-Gaussian noise, identifies maxima of a likelihood statistic which is derived from the SNR and a signal-based 
chi-squared test. This statistic is evaluated over regions of parameter space chosen such that every template is 
\emph{a priori} weighted equally, thus in Gaussian noise there would still be a one-to-one mapping from SNR to 
the expected ranking statistic. 
In contrast, our method accounts for the basic discrepancy between the distributions of signal and noise events 
that will occur, even in Gaussian stationary noise, across \emph{any} coalescing binary search, and allows us to 
derive an optimal single-detector statistic for a given signal distribution.

Section~\ref{sec:results} quantifies how our likelihood-ratio statistic performs on a single-detector search of 
Gaussian data with low-mass, non-spinning binary inspiral signals; we see improvements of tens of percent in the 
number of signals recovered at fixed false alarm rate. We also investigate how the performance of the statistic 
is affected by a discrepancy between the assumed prior over the binary component masses and the true 
distribution of signals. 

Finally in Section~\ref{sec:concl} we discuss possible further developments: extension of the re-weighting 
method to multi-detector searches; possible implementation via machine learning in cases where evaluating the 
signal and noise densities is complex or computationally costly; relations with likelihood-based methods for 
optimizing searches in the presence of non-Gaussian noise transients; and implications of the current highly 
uncertain predictions for coalescing binary source rates \cite{LSCRates}. 

\section{Theoretical motivation} \label{sec:theoretical}
\noindent
In this section we revisit the standard matched filter statistic for detection of a modelled signal in 
Gaussian coloured noise.  We give a condensed overview of the basic strategy of matched filtering as applied to 
compact binary searches; readers less familiar with the background may consult previous descriptions \cite{DuncanThesis,FindChirp} and references therein.  (We also discuss relevant aspects of implementation 
in Section~\ref{sec:findchirp}.)
We point out that when the signal has one or more unknown parameters, 
maximization of the matched filter over those parameters effectively introduces a prior on the parameter
space which is determined by the information metric density, \emph{i.e.}\ the Jeffreys prior, or
equivalently $\sqrt{|\det \Gamma^{ij}|}$ where $\Gamma^{ij}$ is the Fisher matrix. Thus, the matched filter 
is only an optimal statistic for a distribution of signals which matches this prior. We then find a simple
approximation to the optimal detection statistic for a prior distribution which may differ strongly from 
the Jeffreys one.

\subsection{Detection of a signal with unknown parameters}\label{sec:idealized}
Detecting a gravitational wave in data from an interferometric detector can be thought of as a problem of 
model selection \cite{Finn:1992}. The aim is to calculate relative probabilities between the two hypotheses, 
labelled as:
\begin{itemize}
\item[$S$]: There is a signal $h(f)$ and additive Gaussian noise of power spectrum $S(f)$ in the data
\item[$N$]: There is no signal, but only noise with power spectrum $S(f)$ in the data.
\end{itemize}
For a set of observed data $d$ 
we write
\begin{align}
 \frac{P(S|d)}{P(N|d)} &= \frac{P(S)}{P(N)}\frac{p(d|S)}{p(d|N)} \equiv 
 \mathcal{B}_{\rm SN}(d)\frac{P(S)}{P(N)},
\end{align}
where $\mathcal{B}_{\rm SN}(d)$ is the likelihood ratio between the hypotheses (also known as the Bayes 
factor) and $P(S)/P(N)$ is the prior odds in favour of the signal hypothesis. 

For an exactly known signal waveform $h$ with no parameters, and in Gaussian noise, the likelihood ratio 
can be written \cite{Finn:1992,FinnChernoff:1993,CutlerFlanagan:1994}
\begin{align}
\mathcal{B}_{\rm SN}(d)&= \frac{ \exp\left[ -\frac{1}{2}\Braket{d-h|d-h} \right] }
 { \exp\left[ -\frac{1}{2}\Braket{d|d} \right] } 
 = \exp\left[ \Braket{d|h}-\textstyle{\frac{1}{2}}\Braket{h|h} \right] \label{e:Lratio}.
\end{align}
Here $\Braket{\cdot|\cdot}$ is the usual one-sided noise-weighted inner product defined over the space of 
possible data streams $d$; for discrete data in the frequency domain with sampling interval $\Delta f$ we 
have 
\begin{align}
\Braket{a|b} \equiv 4 \Delta f\, \textrm{Re} \left[ \sum_i \frac{{a^*}_i b_i}{S_i} \right],
\end{align}
where $i$ indexes frequency samples from $0$ up to (but not including) the Nyquist frequency.
Since $h$ is constant, $\mathcal{B}_{\rm SN}$ is monotonic in $\Braket{d|h}$; this quantity, called the 
matched filter, is thus an efficient detection statistic~\cite{Finn:1992,DuncanThesis}. 
The Neyman-Pearson lemma states that such a statistic is optimal, yielding the maximum possible detection 
probability at fixed false alarm probability. 

However, in searching for an unknown inspiral signal 
the waveform arriving at the detector has several free parameters. Thus it is necessary to calculate (an 
approximation to) the optimal statistic for a signal with unknown parameters and compare this statistic with 
the standard matched filter. 
We first consider a parametrised signal without assuming any special form, except concerning the signal 
amplitude which we take as unknown and continuously variable between $0$ and infinity. Write the 
signal as 
\begin{equation}\label{hnorm}
 h(\pvec) = \A \hh(\pvec), \qquad \mathrm{where} \qquad \Braket{\hh|\hh} = 1,
\end{equation}
and where $\pvec$ is a set of parameter values in the space $\Theta$, which we consider as a smooth 
manifold; the normalized signal $\hh(\pvec)$ is restricted to a sub-manifold by the condition 
(\ref{hnorm}). 
If considering one specific point in parameter space $\pvec$, the likelihood ratio is 
\begin{equation} \label{likelihood_theta}
 L_{\rm SN}(d;\A,\pvec) = \frac{p(d|S;\A,\pvec)}{p(d|N)} = 
 \exp\left[ \A\Braket{d|\hh(\pvec)} - \frac{\A^2}{2}\Braket{\hh(\pvec)|\hh(\pvec)} \right].
\end{equation}
Then allowing $\A$ to vary, we find a maximum of likelihood at $\A_\ml=\braket{d|\hh(\pvec)}=\rho(\pvec)$, 
using the standard matched filter notation $\rho = \Braket{d|h}/\sqrt{\Braket{h|h}} = \braket{d|\hh}$, and 
the likelihood ratio for the specific parameter values chosen is $p(d|S;\pvec)/p(d|N) = 
\exp\left[\rho(\pvec)^2/2\right]$. 
If we now consider varying the parameters $\pvec$, obviously the point with maximum $\rho(\pvec)$ also 
maximizes $L_{\rm SN}$. 

This straightforward choice does not, though, solve the problem in hand. The marginalized likelihood ratio 
including unknown parameters is 
\begin{equation} \label{eq:margL}
 \mathcal{B}_{\rm SN}(d) = \frac{\int_\Theta d\A\, d\pvec\, p(\A,\pvec|S) p(d|S;\A,\pvec)}{p(d|N)}
 = \int_\Theta d\A\, d\pvec\, p(\A,\pvec|S) L_{\rm SN}(d;\A,\pvec),
\end{equation}
where $p(\A,\pvec|S)$ is a prior distribution for the signal parameters which could arise from our knowledge 
of probable astrophysical sources and their evolution. This expression is also an optimal statistic, in the
Neyman-Pearson sense, for the signal distribution described by the prior.

In general the integral over $\Theta$ must be evaluated numerically (\emph{e.g.}\ using nested 
sampling~\cite{VeitchVecchio:2010} or thermodynamic integration~\cite{LittenbergCornish:2009}),
but under some conditions---essentially, 
the presence of a sufficiently loud signal---it may be approximated via expansion around the maximum which 
at first non-trivial order yields the ``linear signal approximation'' (LSA: 
see for instance \cite{PoissonWill,Vallisneri:2007ev}). This approximation assumes a well-defined peak of 
the likelihood, Eq.~(\ref{likelihood_theta}), for a signal of parameters $\pvec^\ml$ within $\Theta$, near 
which the inner products involving $h(\pvec)$ vary smoothly over the parameter space. 
\cite{CornishLittenberg:2007} finds good agreement between this approximation and a full Bayes factor 
calculation in the high SNR limit when analysing white dwarf binary systems in mock LISA data. 
Due to the exponential dependence of $L_{\rm SN}$ on the matched filter $\rho(\pvec)$ (always assuming 
Gaussian noise) the 
integral over the parameter space is likely to be dominated by a domain near the maximum likelihood point; 
furthermore, although secondary (disconnected) maxima of likelihood will generally occur, their contributions 
to the Bayes factor will in almost all cases be subdominant compared to the global maximum.  Here we 
approximate the result by the integral over a single peak; summing the evidence over possible multiple peaks 
identified by the search would be a refinement to this approximation. 

We make the further approximation that the prior $p(\A,\pvec|S)$ is sufficiently slowly-varying near the peak 
that it may be taken as constant over regions of $\Theta$ that contribute significantly to the integral; this
allows us to take the prior value at the peak $p(\A=\A_\ml,\pvec=\pvec^\ml|S)$ outside the 
integral.\footnote{Under this approximation it is also the case that the posterior distribution $p(\A,\pvec|S)
p(d|S;\A,\pvec)$ varies as the signal likelihood near the maximum.}  
The usefulness of this approximation will clearly depend on the actual behaviour of the prior over such regions, 
whose size and shape are determined by the variation of the likelihood.  Since the likelihood is increasingly 
strongly peaked for higher-amplitude signals, the constant prior approximation will become increasingly good in 
the strong signal limit, which is also where the LSA is valid.  

If this, or other of our approximations, fail badly enough, the ranking statistic we derive will not be (close 
to) optimal for the signal distribution considered, and may not represent an improvement in detection efficiency
over the maximum matched filter $\rho_{\max}$.  We will see this is the case over a relatively small subset of 
the parameter space considered in our example application (Section~\ref{sec:results}): our chosen prior, 
while uniform in component masses, has a singularity at the line $m_1=m_2$ in coordinates for which the 
likelihood is well described by a Gaussian peak of predictable width.  Na\"ively applying the constant prior 
approximation for maxima sufficiently close to this singularity could produce an error of indefinitely large 
size, which could result in random noise fluctuations being wrongly assigned a high ranking value, hurting the 
efficiency of the statistic.  In principle we could go beyond this approximation by expanding the prior about 
the maximum-likelihood point in appropriate coordinates, or by attempting to explicitly carry out the integral 
(\ref{eq:margL}).  However this would go beyond the intention of our study, which is to quantify the 
\emph{leading-order} effects of the signal prior; also, in practice we find that a simple ad-hoc regularization 
of the singularity leads to a statistic which is demonstrably more efficient than $\rho_{\max}$, and we find 
no evidence of any substantial loss in efficiency due to our simple modelling of the prior: see 
Section~\ref{sec:evaluation}. 

The maximum likelihood condition implies 
\begin{align}
\A_\ml &= \Braket{d|\hh^\ml} \equiv \rho_{\max}, \nonumber \\
 \Braket{\partial_i\hh^\ml|d-\hh^\ml} &\equiv 
 \Braket{\frac{\partial\hh}{\partial\theta_i}|d-\hh^\ml}_{|\pvec^\ml} = 0, \label{ML_condition}
\end{align}
and the normalization of the filters $\braket{\hh(\pvec)|\hh(\pvec)}=1$ implies further that
\begin{align}
 \Braket{\hh(\pvec)|\partial_i\hh(\pvec)} &=0, \\
 \Braket{d|\partial_i\hh^\ml} = \left(\frac{\partial \rho}{\partial\theta_i}\right)_{|\pvec^\ml} &=0. 
 \label{ML_condition2}
\end{align}
We expand about the maximum to first derivative order in the waveform $\partial_i\hh$, where now
the parameters indexed by $i$ will not include the amplitude, but keep terms to \emph{second} order in 
parameter deviations $\Dtheta_i = (\theta_i-\theta_i^\ml$).\footnote{Second derivatives of $\hh$ will also 
contribute to this order in $\Dtheta_i$, but at sub-leading order in amplitude $\A_\ml$ (see 
\emph{e.g.}~\cite{PoissonWill}), thus we omit them here.} 
We enforce the norm condition to this order via  
\begin{gather}
 \hh(\pvec^\ml+\Delta\pvec) = C \left[ \hh^\ml + \Dtheta_i\partial_i\hh^\ml + \cdots \right], \nonumber \\
 C^2 \left[ \Braket{\hh^\ml|\hh^\ml} + 2\Dtheta_i\Braket{\hh^\ml|\partial_i\hh^\ml} + 
 \Dtheta_i\Dtheta_j\Braket {\partial_i\hh^\ml|\partial_j\hh^\ml} \right] = 1 \nonumber \\
 \implies C \simeq 1 - \frac{1}{2} \Dtheta_i\Dtheta_j\Braket{\partial_i\hh^\ml|\partial_j\hh^\ml} 
 \equiv 1 - \frac{1}{2} \tilde{g}_{ij}(\pvec^\ml) \Dtheta_i\Dtheta_j,
\end{gather}
where we have pre-emptively defined the parameter space metric $\tilde{g}_{ij}=\braket{\partial_i\hh|
\partial_j\hh}$ (see \cite{Balasubramanian:1995bm,Owen:1996}). For the amplitude we write $\A=\A_\ml+\Delta\A$. 
Expanding inside the exponential in Eq.~(\ref{likelihood_theta}) around the maximum likelihood point and
using the relations (\ref{ML_condition}-\ref{ML_condition2}) we obtain
\begin{align}
 \log L_{\rm SN}(d;\A,\pvec) &\simeq  C \A_\ml (\A_\ml+\Delta\A) - 
 \frac{(\A_\ml+\Delta\A)^2}{2} \simeq \frac{1}{2} \left[ \A_\ml^2\left(1-\tilde{g}_{ij}(\pvec^\ml) 
 \Dtheta_i\Dtheta_j \right) - \Delta\A^2 + \cdots \right]. 
\end{align}
Thus the metric $\tilde{g}_{ij}$ describes the fall-off in log likelihood as a function of the deviation of 
the parameters from their ML values; it is proportional to the Fisher information matrix (at fixed SNR). 
The \emph{match} $M$, \emph{i.e.}\ the overlap of normalized waveforms, used in constructing template banks 
for coalescing binary searches \cite{Owen:1996}, is also described by this metric via 
\begin{align}
 M(\pvec_\ast,\Delta\pvec) = \Braket{\hh(\pvec_\ast)|\hh(\pvec_\ast+\Delta\pvec)} 
 \simeq 1 - \frac{1}{2}\tilde{g}_{ij}(\pvec_\ast) \Dtheta_i\Dtheta_j, 
\end{align}
for a small deviation $\Delta\pvec$ away from $\pvec_\ast$.

The integral in Eq.~(\ref{eq:margL}) may be performed approximately to yield 
\begin{align} \label{Bayes_approx_result}
 \mathcal{B}_{\rm SN}(d) \simeq p(\A=\rho_{\max},\pvec=\pvec^\ml|S) 
 \exp \left( \frac{\rho_{\max}^2}{2} \right) 
 \rho_{\max}^{-m} \sqrt{ \frac{(2\pi)^{m+1}}{\det \tilde{g}(\pvec^\ml)} },
\end{align}
where $m$ is the dimensionality of the parameter space \emph{excluding} the amplitude. 
This expression is no longer monotonic in the maximum matched filter value $\rho_{\max}$ (for $\rho_{\max}>1
$).\footnote{Although this approximate formula for the Bayes factor grows without limit as $\rho_{\max}
\rightarrow 0$, this does \emph{not} indicate an increasing preference for signal over noise in the low-SNR 
limit.  First, the approximation is only valid for (at least moderately) high SNR, certainly not for $\rho
\lesssim 1$; second, in any case the probability that the \emph{maximum} matched filter value over any extended 
region of parameter space is $\lesssim 1$ is vanishingly small.}  Including 
the leading dependence of $\mathcal{B}_{\rm SN}$ on the parameters of the maximum has two effects: firstly via
the metric density $\sqrt{\det\tilde{g}}$, and secondly by allowing a prior over the parameter space. Note that 
the ratio of the prior density to the metric density is invariant under a change of coordinates over $\Theta$. 

The approximate Bayes factor of Eq.~(\ref{Bayes_approx_result}) is only monotonic in $\rho_{\max}$ if the 
signal prior is numerically equal to the metric density. Thus, \emph{maximizing the matched filter over a
parameter space implicitly introduces a prior over the space: the maximum matched filter is not an 
optimal statistic unless the signal distribution is equal to the metric density.}  The implicit prior 
is given by 
$ p_{\rm imp}(\A,\theta|S) = f(\A) \sqrt{\det \tilde{g}(\theta)} $,
where $f(\A)$ is a smooth function of signal amplitude such that $\mathcal{B}_{\rm SN}$ is a monotonically
increasing function of $\rho_{\max}$ at $\A=\rho_{\max}$ ($\rho_{\max}>1$).

A similar implicit prior has been investigated in the case of amplitude parameters of continuous gravitational 
waves from isolated neutron stars, where it was indeed found that accounting for the correct prior could 
lead to improvements in detection efficiency over the maximized matched-filter (``F-statistic'') method 
\cite{Prix:2009}. The use of explicit, physically motivated priors has also been shown to improve the 
efficiency of searches for unmodelled (or, to be more accurate, weakly-modelled) GW bursts 
\cite{Searle:2008ap} compared to standard analysis methods. 
Note that an optimisation of template bank placement using a prior signal distribution was proposed in 
\cite{Roever:2010}, but under a criterion of maximum efficiency at fixed computing cost (taken as 
proportional to the number of templates) rather than at fixed false alarm rate.

\subsection{Search parameters of coalescing binary signals}\label{sec:inspiralparams}

We now explore the implications of this result for binary coalescence signals by considering their 
parameters as seen, for simplicity, by a single interferometric detector. (Multi-detector searches will
be discussed in section~\ref{sec:concl} as a topic of future development.)
The gravitational wave signal from a binary inspiral 
is dependent on up to 15 parameters, in the case of a binary with a quasi-circular orbit and spinning 
components (neglecting matter effects).
We write the signal in the form $h(f)=\A \hh(f;\mvec) \exp(-2\pi i f t_c - i\coa)$, where the ``intrinsic 
parameters'' $\mvec$ contain the masses and the spin vectors of the binary components, separating out the 
dependence on the coalescence time of the signal $t_c$ and coalescence phase $\coa$.  For dominant-mode signals 
in ground-based detectors for which the detector antenna pattern is approximately constant over their duration, 
the phase evolution $\hh(f;\mvec)$ is a function of intrinsic parameters only, whereas the amplitude $\A(\mvec,
\ra,\dec,\psi,\iota,d_L)$ may depend on the intrinsic parameters and on the inclination $\iota$ angle of the 
binary, the polarisation angle $\psi$, the luminosity distance $d_L$, and the position on the sky relative to 
the detector $\ra,\dec$. 
Together, these parametrise the signal hypothesis $\{\mvec,\A,t_c,\coa\}\in\Theta$. 

\subsection{Coalescence phase and the coincidence metric}

As described in \cite{DuncanThesis,FindChirp}, in searching for binaries where the orbital plane does not
precess, the coalescence phase $\coa$ is not explicitly searched over 
in matched filtering single-detector data. Rather, a complex matched filter is evaluated via Fast Fourier 
Transform and the analogue of the matched filter for known signals is found to be the absolute value of the 
complex filter. For a single filter template (a point in parameter space $\mvec$) the likelihood ratio 
marginalized over $\coa$ with uniform prior was shown to be monotonic in the ``two-phase'' SNR $\rho = 
|z|/\sigma$, where $z = \braket{d|h(\mvec,t_c,\coa=0)}+i\braket{d|h(\mvec,t_c,\coa=\pi/2)}$ and $\sigma$ is 
a template normalization factor; $\rho$ is thus an optimal statistic for a signal of unknown phase. 
The maximum $\rho$ value is also equal to the maximum-likelihood amplitude $\A_\ml$.

Since the dependence of the signal on $\coa$ is simple and exactly known, we may similarly marginalize the 
likelihood ratio over $\coa$ before expanding in the amplitude and other parameters; for a detailed 
calculation refer to appendix~\ref{app:coa}.  This is expected to be more accurate than a quadratic 
expansion around the maximum likelihood value; we explicitly show, though, that the two methods agree for 
high-amplitude signals. The phase-marginalized likelihood ratio is then
\begin{align}
 L_{\rm SN}'(d;\A,\pvec) &\equiv \int \frac{d\coa}{2\pi}\, L_{\rm SN}(d;\coa,\A,\pvec) \nonumber  
 \simeq \exp\left(-\frac{\A^2}{2}\right)
 I_0\left[ \A\sqrt{\braket{\hh_0(\pvec)|d}^2+\braket{\hh_{\pi/2}(\pvec)|d}^2} \right],
\end{align}
where $I_0$ is the modified Bessel function of the first kind; this may be expanded in the other parameters
and marginalized to yield 
\begin{align} 
 \int d\A\,d\pvec\, p(\A,\pvec|S) L' & 
 \simeq p(\A=\A_\ml,\pvec=\pvec^\ml|S) L_\ml' \int d(\Delta \A)\,e^{-\Delta\A^2/2} 
 \int d^N\pvec \exp\left[-\frac{\A_\ml^2}{2} \gamma_{pq}\Delta\theta_p\Delta\theta_q\right] \\
 &\approx p(\A=\rho_{\max},\pvec=\pvec^\ml|S) \exp\left(-\frac{\rho_{\max}^2}{2}\right)I_0\left(\rho_{\max}^2\right)
 \rho_{\max}^{-n} \sqrt{\frac{(2\pi)^{n+1}}{\det \gamma(\pvec^\ml)}},
\label{Bayes_with_phase}
\end{align}
where as before we identify $\A_\ml$ with the local maximum matched filter $\rho_{\max}$; $p,q$ label the 
$n$ parameters other than $\coa$ and $\A$; and $\gamma_{pq}$ is the phase-projected metric
\begin{align}\label{gammapq} 
 \gamma_{pq} = \tilde{g}_{pq} - \frac{\tilde{g}_{p\coa}\tilde{g}_{q\coa}}{\tilde{g}_{\coa\coa}}
 = \left[ \Braket{\partial_p\hh_0|\partial_q\hh_0} - \Braket{\partial_p\hh_0|\hh_{\pi/2}}
 \Braket{\partial_q\hh_0|\hh_{\pi/2}} \right].
\end{align} 
This metric was used in \cite{Robinson:2008un} to describe the dependence of the phase-maximized 
likelihood on the remaining signal parameters, \emph{i.e.}\ coalescence time and component masses. Its 
inverse gives the expected error matrix over these parameters for the maximum-likelihood point, compared 
to the true values when a signal is present; it was therefore used to implement a consistency test for 
``triggers'' recorded at maxima of the matched filter $\rho$ in the data streams from different detectors
\cite{Robinson:2008un}.

\subsection{Prior parameter distribution}

In order to evaluate an optimized statistic we need to specify the prior $p(\A,\mvec,t_c,\coa|S)$. The 
coalescence phase and time dimensions are trivial and have uniform priors; also, at least for stationary
detector noise, the distributions over other parameters do not depend on the values of $t_c$ and $\coa$.
\footnote{However, if the detector sensitivity changes with time the prior distribution over $t_c$ of signals 
with a given expected SNR, i.e.\ a given value of $\A$, is \emph{not} uniform. We will not consider that case 
here.} 
Further, for a distribution of coalescing binary sources uniform in physical volume, and where there are no 
correlations between the luminosity distance and any other physical parameter, one may show (see for instance 
\cite{Schutz:2011tw}) that there is a universal distribution of signals over the amplitude $\A$, 
$p(\A|S) \propto \A^{-4}$, thus the prior may be factorized as 
\begin{align}
 p(\A,\mvec,t_c,\coa|S) \propto
p(\A|S)p(\mvec|S,\A) \propto \A^{-4}p(\mvec|S,\A),
\end{align}
where $p(\mvec|S,\A)$ is the distribution over the intrinsic parameters spanning the template bank at constant 
expected SNR. We will discuss this prior distribution further in Section~\ref{triggerstats} where it will 
appear as the rate distribution of signals of a given SNR over the search parameter space. 

\section{Implementation in matched filter search}\label{sec:findchirp}
\noindent
In this section we describe some currently used methods for searching a parameter space of coalescing 
binary systems and show how the output of such a search corresponds to the statistical derivation of the 
previous section. We are thus able to implement our approximately optimal statistic using the ``triggers'' 
produced by standard search algorithms \cite{FindChirp} in combination with a parameter-dependent 
re-weighting. 

\subsection{Template bank layout}

To begin, the template bank, a set of discrete parameter values $\mvec_k$, is chosen in order to minimize 
the maximum possible ``mismatch'' between a signal waveform and the best fitting template. The complex 
matched filter is evaluated for each of the templates, and maxima of its modulus $\rho$ over the bank, and 
over short time intervals, are recorded. 
For dominant-mode signals from non-precessing binary systems, the amplitude, coalescence time and 
coalescence phase can be efficiently searched by a single Fourier transform, leaving the two binary masses 
and possible non-precessing spin components as dimensions to be covered by the bank; we will refer to 
these as ``intrinsic parameters''. 

Each template ``covers'' a finite region of the intrinsic parameter space, i.e.\ has a sufficiently high 
match $M$ with signals in the region. The match, for a signal with intrinsic parameters $\mvec_S$, is
the ratio of the expected SNR recovered by a template at $\mvec_S+\delta\mvec$, after maximizing over 
time and phase, to the optimal SNR recovered by a filter with the signal's parameters $\mvec_S$. 
For broad-band detectors, the total region where astrophysical signals may be present is much larger 
than that covered by a single template, therefore we require a ``bank'' of many templates; 
these are typically laid out in a grid designed to achieve a specific minimal match $M_{\min}$ with at 
least one template for any signal in the total region (in recent searches 
$M_{\min}=0.97$)~\cite{Owen:1996,OwenSathya:1999,Babak:2006ty,Cokelaer:2007}.

The template bank is created with the help of a $N$-dimensional metric $g(\mvec)$ over the intrinsic 
parameter space,\footnote{Not to be confused with the general parameter metric $\tilde{g}$ considered earlier!} 
which is simply a projection along $t_c$ of the coincidence test metric:
\begin{align}
 g_{kl} = \gamma_{kl} - \frac{\gamma_{{t_c}k}\gamma_{{t_c}l}}{\gamma_{{t_c} {t_c}}}.
\end{align}
As a result of this construction, the density of an ideal template bank over the parameter space is 
proportional to the metric density \cite{Owen:1996,OwenSathya:1999} (in some coordinates for $\mvec$):
\begin{align} 
 n_T(\mvec) = \mathrm{const.}\times (1-M_{\min})^{-N/2} \sqrt{\det g(\mvec)}.
\end{align}
In practice, geometrically constructed banks (as described in \cite{Babak:2006ty,Cokelaer:2007}) for 
non-spinning binaries may make use of the near-constancy of metric components in coordinates $\tau_0$, 
$\tau_3$ defined as 
\begin{align}
 \tau_0=\frac{A_0}{\eta}(m_1+m_2)^{-5/3}, \qquad \tau_3=\frac{A_3}{\eta}(m_1+m_2)^{-2/3}, 
\end{align}
where $\eta\equiv m_1m_2/(m_1+m_2)^2$ and $A_0,A_3$ are numerical constants. Then a regular lattice with 
fixed spacing and orientation over $\tau_0,\tau_3$, implicitly using a metric that is constant and flat 
over this space, yields a bank for which the deviation from the theoretically correct, fixed minimal match 
$M_{\min}$ over the whole parameter space is relatively small for low-mass (non-spinning) binary systems 
\cite{Keppel:2013kia}. The hexagonal bank placement used for recent searches of LIGO and Virgo data (see 
\emph{e.g.}~\cite{S5Year1,Colaboration:2011np,ihope}) uses a refinement of the regular lattice via an 
adaptive algorithm that locally adjusts the grid spacing and orientation to account for gradual
variation of the metric components over $\tau_0,\tau_3$ space~\cite{Cokelaer:2007}. 

\subsection{Matched filter triggers and clustering}

Each template $\mvec_k$ is used to filter the data stream from a detector $d(t)$ producing a time series 
$\rho_k(t_c)$ of the complex matched filter modulus. In Gaussian noise, all template time series have the 
same distribution \cite{DuncanThesis}, $p(\rho_k(t_c)) = \rho_k e^{-\rho_k^2/2}$. 
These matched filter values show strong correlations over nearby coalescence times and between nearby 
templates; the filter correlations over small changes in $t_c$ and $\mvec$ can be described by the 
$(N+1)$-dimensional ``coincidence'' metric $\gamma(t_c,\mvec)$.  

We would like to obtain an optimal statistic to distinguish noise events from signals, where events are 
considered to be localized in time and over template intrinsic parameters. In order to use the familiar 
statistics of false alarms, false dismissals, etc., we will model candidate signals and false alarms by 
time-localized events (approximately) obeying Poisson statistics. A process of \emph{clustering} relates
the set of matched filter time series to such localized events (``triggers''). 

\paragraph*{Clustering over time} A signal is expected to give rise to a maximum of the matched 
filter and also to large values in neighbouring time samples, dying away at other times according to the 
template's autocorrelation. 
We wish to allow for multiple signals (or noise events) over a search time, therefore we record local 
maxima of the matched filter at intervals shorter than the expected waiting time between signals. 
These maxima are called \emph{triggers} and will be notated\footnote{The matched filter value
$\rho$ should be understood as a trigger value rather than a time series unless a dependence on time 
$\rho(t_c)$ is explicitly given.} as $\{t_c,\rho,\mvec_k\}$. 
In practice, the maximization window is set to 1\,s or longer, up to the length of the template, in 
order to reduce computational storage requirements, and triggers are also not
recorded if $\rho(t_c)$ does not exceed a predetermined threshold, typically $\rho_{\rm th}=5.5$. 
For a single template, triggers may be modelled as Poisson events with a $\rho$-dependent rate 
$\lambda_1(\rho)$, up to the condition that no more than one trigger may occur within a short time window; 
the approximation becomes increasingly good for smaller $\lambda_1$ (and thus increasingly large $\rho$). 

\paragraph*{Clustering over the template bank} 
Any given high-SNR signal or loud noise fluctuation will still produce strongly-correlated maxima above 
threshold in nearby templates and at similar times. We may again consider recording only the maximum value 
of $\rho(\mvec_k)$ over a correlated region of template parameters and time; 
however, this procedure implicitly assumes a uniform prior signal distribution \emph{over different 
templates}. In other words, maximizing $\rho$ over templates can only yield an optimal statistic if we are 
equally likely to see a signal in each template during any given stretch of data. 

Since we intend to investigate the role played by the signal distribution in the search, if we were to 
cluster by maximizing $\rho$ over the entire bank and over substantial periods of time we might obliterate 
the effect. Nevertheless, over \emph{small} regions of parameter space encompassing, say, order(10) 
templates (a typical full bank contains many thousands) we may consider the prior probability of signal as 
approximately constant between templates, and certainly as varying much less rapidly than the likelihood 
ratio $L_{\rm SN}\sim\exp(\rho^2/2)$. Hence we use the ``trigscan'' algorithm \cite{TrigScan} to cluster by 
recording maxima of $\rho$ locally over template parameters and time, proximity being tested via the 
$(N+1)$-dimensional overlap metric $\gamma(t_c,\mvec)$. 
Triggers widely separated in parameter space (whether occurring at the same time or different times) 
will not affect one another through clustering and can be treated as independent. 

Note also that if the bank has regions where the density of templates is larger than the theoretical 
value required for the desired minimal match, i.e.\ areas of over-coverage, for example due to edge effects, 
then the number of \emph{unclustered} triggers with a given $\rho$ will also show an excess in these regions;
however, the behaviour of triggers \emph{after} clustering should not be affected by such over-coverage.

\subsection{Trigger statistics and an optimized event ranking} \label{triggerstats}

Given a set of clustered triggers, the problem remaining is to find a detection statistic 
$\Lambda(t_c,\rho,\mvec)$ which will maximize the efficiency of detection at a given false alarm rate. 
Triggers will be ranked by $\Lambda$ value and those above a constant threshold $\Lambda_t$ will be 
considered as detection candidates. The false alarm rate is the total number of triggers that exceed 
$\Lambda_t$, divided by the analysis time, for data containing no signals; the detection efficiency, for a 
finite population of signals with a given distribution over physical parameters, is the fraction of signals 
for which an associated trigger can be identified whose statistic exceeds $\Lambda_t$. 
A non-trivial dependence on $t_c$ will be appropriate if the detector sensitivity varies with time or if 
some times are more likely to yield signals than others; here we consider a case where the prior over 
$t_c$ is constant, thus $\Lambda=\Lambda(\rho,\mvec)$. 

Note that the ranking is determined by comparing statistic values calculated for \emph{single} events or
triggers, a choice which corresponds to the procedure of approximating the Bayes factor by an integral over 
a single maximum of likelihood in the previous derivations.\footnote{To go beyond this assumption one would 
evaluate a detection statistic over sets containing multiple triggers, with each set representing the result 
of a search over an extended data set.} 

The Neyman-Pearson optimal ranking is then \cite{Biswas:2012tv}
\begin{align}\label{Lambdaopt}
 \Lambda_{\rm opt}(\rho,\mvec) = \frac{\lambda_S(\rho,\mvec)}{\lambda_N(\rho,\mvec)}
\end{align}
where $\lambda_S$, $\lambda_N$ are Poisson rate densities of clustered signal and noise triggers over 
$\rho$ and over the template parameters. To relate this to the familiar ratio of likelihoods, consider a 
short period of data over which the expected number of signal or noise triggers is small ($\ll 1$), and 
suppose we find one trigger in that time with parameters $\rho,\mvec_k$: the likelihood ratio 
$p(1;\rho,\mvec_k|S)/p(1;\rho,\mvec_k|N)$, which is an optimal statistic for any one trigger, is then 
identical to Eq.~(\ref{Lambdaopt}).\footnote{The probabilities for a single template $\mvec_k$ are 
related to the densities $\lambda_{S,N}(\mvec)$ over the bank by a change of coordinates which cancels in 
this ratio.} 
Although we may not be able to determine the absolute rates of clustered triggers $\lambda_{S,N}
(\rho,\mvec)$ theoretically or from prior information, we can constrain their functional forms. 

\paragraph*{Modelling noise triggers} 
In Gaussian noise, each template produces a predictable distribution of matched filter time samples 
$p(\rho_k(t_c))=\rho_k e^{-\rho_k^2/2}$. However, we are concerned with the distribution of 
\emph{independent} (clustered) noise events. The clustering procedure will alter the distribution of 
trigger $\rho$ values to a function $p(\rho|N)$, which we will model empirically. 
The \emph{rate} of noise triggers with given $\rho$ may also vary over the parameter space. Since the 
correlations between triggers are described by the metric $\gamma(t_c,\mvec)$, which is also used to 
perform clustering, we expect that the noise trigger rate density will be given by the metric density in 
some coordinates over $\mvec$, thus: 
\begin{align}
 \lambda_N(\rho,\mvec) \propto p(\rho|N) \sqrt{\det \gamma(t_c,\mvec)}.
\end{align}
We note in passing that the variation of the metric density $\sqrt{\det\gamma}$ over the 2-dimensional 
space of non-spinning component masses $m_1,m_2$, for the ``low-mass'' space $1\leq m_{1,2} \leq 24$ 
$m_1+m_2\leq 25$, is similar to that of the bank metric $\sqrt{\det g(m_1,m_2)}$. 
Thus we expect that regions containing equal numbers of templates will yield approximately equal numbers of 
clustered noise triggers; 
we were able to check this explicitly for an example search (see Section \ref{sec:results}). 
Small variations in the per-template rate of clustered noise triggers may be due to the variation of the 
$t_c$-components of $\gamma$, representing the filter autocorrelations \cite{Keppel:2013kia}. 

\paragraph*{Modelling triggers from an astrophysical signal population} 
The rate of signal triggers expected depends on both the astrophysical coalescence rates and a number of 
other factors that determine the analysis sensitivity at given SNR (see for instance 
\cite{Fairhurst:2007qj}). We start with
an astrophysical rate density $R(\mvec_S)$, a function of the source intrinsic parameters $\mvec_S$, 
measured in coalescences per year, per cubic Mpc, per volume of parameter space $d^N \mvec_S$. We assume
that coalescences are uniformly distributed over time and space out to the limits of the detector's maximum
sensitive distance, and neglect any effects of redshift. 
The expected SNR of a signal $\bar{\rho}$ is affected by the source's intrinsic parameters (mass and spin), 
angular extrinsic parameters (orientation and sky direction relative to the detector), the distance to the 
detector, and the detector's sensitivity, i.e.\ its noise power spectral density (PSD). 

For a non-spinning inspiral signal where the template's upper frequency cut-off is above the detector 
sensitive band, we have $\bar{\rho} \propto \Mc^{5/6}/D$, where $D$ is the distance to the source and the 
chirp mass is defined as $\Mc \equiv (m_1m_2)^{3/5}(m_1+m_2)^{-1/5}$. The proportionality depends on 
angular parameters and on the detector sensitivity, but this dependence is universal and thus factors out 
of the relative rates of signals with different $(\bar{\rho},\mvec_S)$. 

Formally, considering events within an element of the space of angular extrinsic parameters, but 
explicitly allowing for a varying distance $D$, we have a number density of signals at the detector 
\begin{align}
 dN \equiv \mu_S(\bar{\rho}, \mvec_S)\, dt\, d\bar{\rho}\, d^N\mvec_S = 
 \mathrm{const.}\times R(\mvec_S)\, dt\, dV(\bar{\rho},\mvec_S)\, d^N\mvec_S
\end{align}
where $\mu_S$ is a rate density over the expected SNR $\bar{\rho}$ and source intrinsic parameters 
$\mvec_S$, and $V(\bar{\rho},\mvec_S)$ is the volume of space over which a signal with the given 
parameters would have an expected SNR greater than $\bar{\rho}$. 
Since $V=\mathrm{const.}\times D^3 = \mathrm{const.}\times\Mc^{5/2} \bar{\rho}^{-3}$, we find 
\begin{align}
 \mu_S(\bar{\rho},\mvec_S) = \mathrm{const.}\times R(\mvec_S) \frac{\Mc_S^{5/2}}{\bar{\rho}^4}.
\end{align}
This relation holds with some constant of proportionality for any value of angular parameters, 
therefore it will also hold for the distribution as a whole provided that the distribution over 
angular parameters does not depend on $D$ or $\mvec_S$. 

For signals where the expected SNR does not scale with the chirp mass as above (\emph{e.g.}\ when late 
inspiral or merger lies within the detector's sensitive band), $\Mc_S^{5/2}$ may be replaced by 
$D_{\rm hor}(\mvec_S)^3$, where $D_{\rm hor}$ is the horizon distance at which a coalescence with optimal 
orientation overhead from the detector would have a given, fixed SNR (usually taken as $\bar{\rho}=8$).

Now consider the distribution of signal triggers $\lambda_S(\rho,\mvec)$ over the \emph{recovered} maximum 
matched filter $\rho$ and parameter values $\mvec$. 
For a given signal $\rho$ will not be equal to the expected SNR, nor will the ML parameters be equal to 
those of the source. This parameter inaccuracy is in general caused both by random noise, and by bias in the
template waveform model which may cause the best match to occur at recovered parameter values substantially 
different from the true physical ones; see \cite{BIOPS,Nitz:2013} for relevant discussions of PN inspiral 
waveform systematics. 

In general one should perform a Monte Carlo study to determine the relation between $\mu_S$ and $\lambda_S$, 
but we note that, for ``faithful'' templates with relatively small parameter bias, if the signal distribution 
$\mu_S$ has sufficiently broad support and is slowly-varying, 
then the difference between the two distributions is unlikely to be substantial except at low $\bar{\rho}$ 
where the recovered trigger distribution will obviously be cut off by the threshold imposed in the 
analysis. Hence we may take
\begin{align} \label{signal_rate_approx}
 \lambda_S(\rho, \mvec)  \begin{cases} 
   \; \simeq \mu_S(\bar{\rho}=\rho, \mvec_S=\mvec), & \bar{\rho} > \rho_{\rm th} \\
   \; = 0, & \bar{\rho} < \rho_{\rm th}.
\end{cases}
\end{align}
If $\mu_S(\bar{\rho},\mvec_S)$ does vary strongly over its arguments, for example being strongly peaked 
over mass or with sharp cutoffs inside the search parameter space, the errors in recovered parameters must 
be treated more rigorously. The ``slowly-varying'' condition for Eq.~(\ref{signal_rate_approx}) to be valid 
is analogous to that for neglecting the variation of the prior density when marginalizing over a peak of 
likelihood.

We then find the approximately optimized statistic, for the source population described by $R(\mvec_S)$: 
\begin{align} \label{eq:Lambdaopt}
 \Lambda_{\rm opt}(\rho,\mvec_k) \simeq \frac{1}{\rho^4 p(\rho|N)}
 \frac{R(\mvec_S=\mvec_k)D_{\rm hor}(\mvec_k)^{3}}{\sqrt{\det \gamma}_{|\mvec_k}}, 
\end{align}
recalling that $\mvec_k$ is the parameters of the $k$'th template. 

Although this statistic is the ratio of rate densities over the parameters of clustered triggers, whereas 
the Bayes factor evaluated approximately in Eq.~(\ref{Bayes_approx_result}),~(\ref{Bayes_with_phase}) is a 
ratio of marginalized likelihoods for the detector time series, the two are identical in form up to a 
constant numerical factor if we identify the prior signal distribution with $R(\mvec)D_{\rm hor}^3/\rho^4$ 
and the remaining factors with the inverse of the noise trigger density. This indicates we are solving
essentially the same optimization problem in both cases given our approximations and assumptions; 
specifically, that we consider sufficiently high maxima of likelihood (at least above the threshold for 
trigger generation), and we marginalize or cluster triggers over regions of parameter space containing no 
more than one such maximum.

\subsection{Comparison with SNR ranking}

The optimized statistic for a given signal population in Gaussian noise differs in three respects from a 
ranking of events on matched filter SNR $\rho$. First, via the astrophysical rate distribution; second, via 
the dependence of sensitive volume on the signal parameters; third, via the expected rate density of noise 
triggers. If, as would be the standard choice in Gaussian noise, we were to rank events via $\rho$ alone, 
our search would still be optimized for \emph{some} signal population, specifically that described by 
\[
 R_{\rho}(\mvec_S) = \sqrt{\det\gamma}_{|\mvec_S} \cdot D_{\rm hor}(\mvec_S)^{-3}.
\]
It may be plausible that the astrophysical coalescence rate varies as some negative power of the horizon
distance, at least within some restricted space of masses, 
but there is no physical reason why we would expect it to vary proportionally to the metric density 
associated with the search filters, which may depend on both the detector noise PSD and the signal 
model used. 
Furthermore, many different 
searches may be performed which will generally have different values of $\gamma(t_c,\mvec)$, whereas 
the astrophysical population will have a fixed and constant distribution $R(\mvec_S)$. 

\begin{figure}
\includegraphics[width=0.6\textwidth]{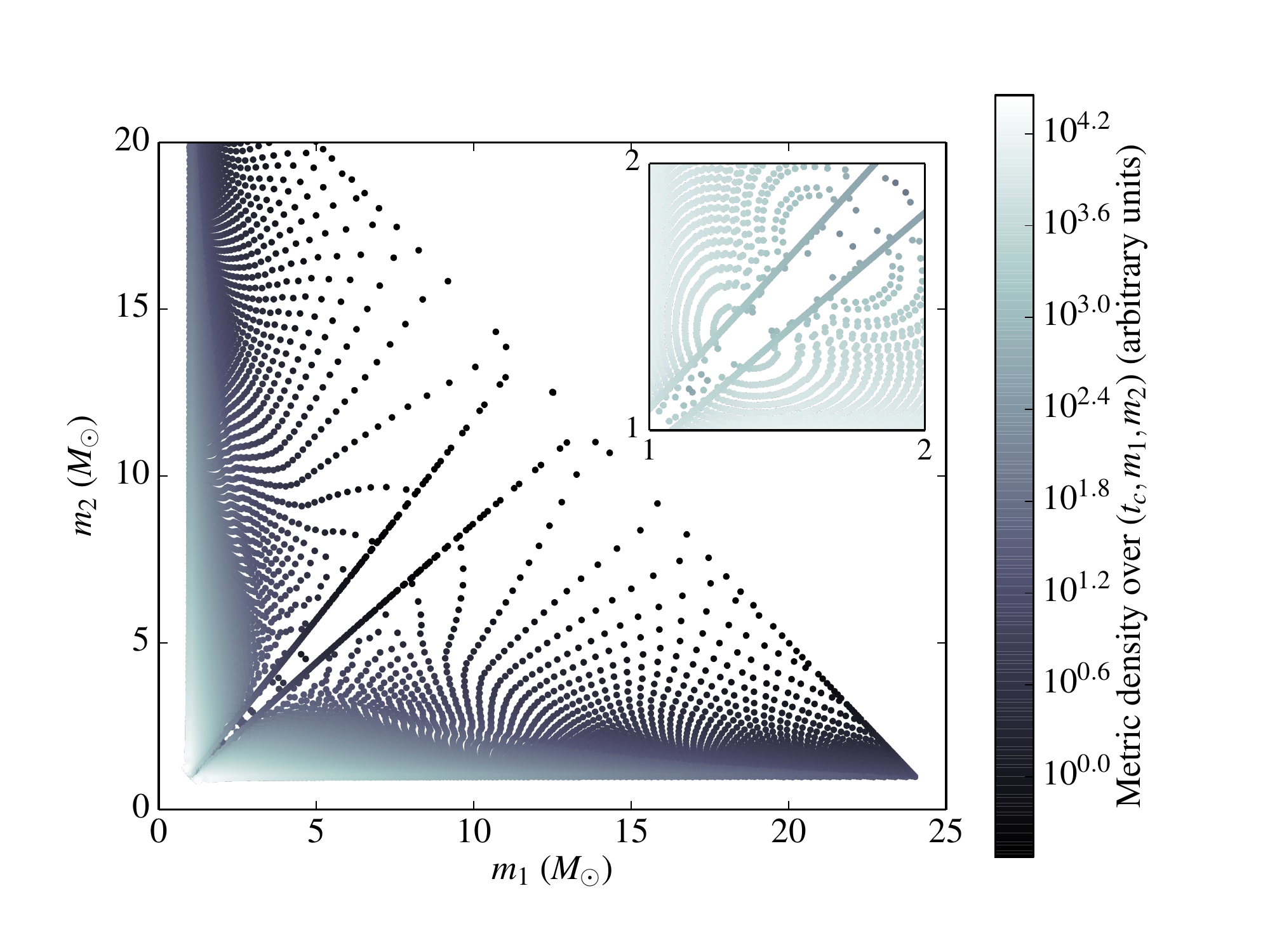}
\vspace*{-0.6cm}
\caption{Metric density calculated by the standard LAL template bank code at each $(m_1,m_2)$ point in the 
``low-mass'' bank, transformed to $(t_c,m_1,m_2)$ coordinates, and regularized as described in 
Section~\ref{sec:evaluation} by adding a small constant (equal to $0.3$, in the units shown) in quadrature.  
The apparent excess density of templates near $m_1=m_2$ is necessary to avoid potential under-coverage at 
high $\eta$.
\label{fig:lowmass_density}
}
\end{figure}

Still, if the variation in $\sqrt{\det\gamma}$ over the search space were 
negligibly small compared to variations in likelihood due to differences in $\rho$, then the matched filter 
value might still be a good approximation to an optimal statistic. We show, though, that this is \emph{not} 
the case for the standard ``low-mass'' inspiral search space in Fig.~\ref{fig:lowmass_density}, where the 
metric density evaluated by a standard LIGO Algorithm Library (LAL)~\cite{LALSuite} bank placement code 
at 2PN order, transformed to $t_c,m_1,m_2$ coordinates, is plotted for templates with component masses 
between $1$--$24$\,\Msun\ and total mass $2$--$25$\,\Msun, for a lower frequency cutoff of 40\,Hz and a 
noise PSD representative of the ``early advanced LIGO'' sensitivity range given in \cite{Aasi:2013wya}. 
There is a variation of several orders of magnitude between the sparsest region and the densest near 
$m_1=1.5$, $m_2=1$. 
Hence, we may expect a significant difference between the ranking of events by an optimized statistic
$\Lambda_{\rm opt}$ compared to ranking by $\rho$. 

\section{Example application: search for low-mass non-spinning binary coalescence}\label{sec:results}
\noindent
In this section we will quantify the differences in search efficiency produced by using the optimized 
statistic described in the previous section for a simple example of a search parameter space and possible 
signal distributions over it. 

\subsection{Simulated data and trigger generation}

The space of ``low-mass'' binaries with non-spinning components of mass between $1$--$24$\,\Msun\ and total 
mass up to $25$\,\Msun was the basis of a search 
in recent LIGO-Virgo data~\cite{Colaboration:2011np}. 
Using standard LAL codes we generated template banks at 2PN order for this search space, and obtained noise 
background triggers by matched filtering approximately 18 days of Gaussian coloured noise with a sensitivity 
curve (PSD) representative of the ``Early Advanced LIGO'' operational sensitivity shown in \cite{Aasi:2013wya},
for the LIGO Hanford (LHO) and LIGO Livingston (LLO) detectors. 
We used the stationary-phase approximation (SPA) filters of \cite{FindChirp}, corresponding to the TaylorF2 
approximant, with a lower frequency cutoff of 40\,Hz, and approximately 11300 templates were placed for each
detector.  The 2PN order was chosen for simplicity given that the appropriate metric \cite{OwenSathya:1999} and 
template placement algorithm are well-established; metric-based template banks for (hopefully) more accurate 
waveform models are just becoming available \cite{Brown:2012qf,Harry:2013tca}. 

For simplicity we simulated an astrophysical distribution of signal events with the property that the product 
$R(m_1,m_2)\Mc^{5/2}$ is constant over the search space, implying a constant rate density of signals at fixed 
SNR per $dm_1\,dm_2$ element. We note that this distribution implies an astrophysical rate which is a 
\emph{negative} power of mass; this is consistent with the simple expectation that the distribution of compact 
objects may mirror a stellar initial mass function up to some maximum mass.
The signal distribution was implemented by generating a set of simulated signals with the TaylorT2 approximant
\cite{DrozSPA:1999,Nitz:2013} at 2PN order with random 
parameters distributed uniformly over the component masses and over angular parameters, then rescaling the 
distances to achieve a $1/\bar{\rho}^4$ distribution of expected SNRs, above a minimum value chosen below the 
search threshold $\rho_{\rm th}=5.5$, and up to a maximum $\bar{\rho}$ value well above any noise 
event\footnote{More precisely, the network SNR for the LHO and LLO detectors 
$(\bar{\rho}_H^2+\bar{\rho}_L^2)^{1/2}$ was rescaled to the target $1/\bar{\rho}^4$ distribution up to a 
maximum of 20; virtually all the loudest signals in this distribution are well above any (Gaussian) noise event 
in either detector.}. 

To reduce the number of random noise triggers in our signal event sample we also imposed the parameter
consistency test of~\cite{Robinson:2008un} between triggers in the two LIGO detectors. Despite the application 
of clustering, there may occasionally be more than one coincident trigger recovered near the coalescence time 
of a simulated signal; in these cases we took the trigger with the highest $\rho$ within $\pm0.025$\,s of that 
time to represent the signal.\footnote{This choice may be sub-optimal when using a ranking statistic other 
than $\rho$; we should instead use the trigger with the best statistic value associated with the simulated 
signal. Doing so would only strengthen any gain in efficiency due to an improved statistic.} 
We thus obtained approximately 8000 simulated signal events. 

Since our signal recovery involved choosing the loudest trigger by SNR over an $0.05$\,s window around the
simulated signal time, to treat noise triggers consistently we also clustered them over $0.05$\,s time windows. 
This extra clustering step had a negligible effect on triggers above a SNR of approximately $6.0$ and we found 
our results were not affected by it except at very high false alarm rates (FARs).  We obtained approximately 
$5.2\times10^6$ clustered noise triggers per detector. 

\subsection{Evaluation of the optimized statistic}
\label{sec:evaluation}
Then, considering only triggers from a single detector's data, we evaluated the optimized statistic given 
by the RHS of Eq.~(\ref{eq:Lambdaopt}) for both background and simulated signal triggers, using $m_1,m_2$ 
coordinates for the signal rate and metric. We modelled the noise trigger distribution $p(\rho|N)$ above
threshold by a falling exponential $\propto e^{-\alpha\rho}$, finding $\alpha\simeq\ln(250)$. 
The metric components are pre-computed to 2PN order
in $(t_c,\tau_0,\tau_3)$ coordinates; the density in $(t_c,m_1,m_2)$ space is then given by 
$\sqrt{\det\gamma(t_c,m_1,m_2)} \equiv |\det J|\sqrt{\det \gamma(t_c,\tau_0,\tau_3)}$, where $J$ is the 
Jacobian with
\begin{align} \label{detJ_m1m2_tau0tau3}
 \det J = \det \begin{bmatrix} \frac{\partial \tau_0}{\partial m_1} & \frac{\partial\tau_3}{\partial m_1} \\ 
 \frac{\partial \tau_0}{\partial m_2} & \frac{\partial \tau_3}{\partial m_2} \end{bmatrix}
 = \frac{A_0A_3}{m_1^3m_2^3} (m_1+m_2)^{2/3} (m_1-m_2),
\end{align}
where we omitted trivial elements involving $t_c$ (unity on the diagonal and zero elsewhere). 

\paragraph*{Regularization and edge effects} The resulting density $\sqrt{\det \gamma(t_c,m_1,m_2)}$ vanishes 
on the $m_1=m_2$ line, which would na\"\i vely assign an \emph{infinite} statistic value to any triggers in 
equal-mass templates.  This singularity presages
the breakdown of our approximation of a constant prior 
over a peak of likelihood near this line.  The width of such a peak, at a given SNR, is approximately constant in 
$(\tau_0,\tau_3)$ coordinates; in these coordinates the prior density is multiplied by the inverse of the
Jacobian $J$, which blows up at $m_1=m_2$.\footnote{Equivalently, in $(m_1,m_2)$ coordinates the prior is 
constant but the \emph{area} of integration in parameter space covered by a peak of likelihood, at least in the
linear signal approximation, blows up at equal mass.}  
Thus, for SNR maxima close to this line the appropriate optimal statistic is not given by 
Eq.~(\ref{Bayes_with_phase}), rather it would require an explicit integration with a varying prior, for which the 
result would not diverge. 

In addition to this singularity we will in general meet artefacts due to the discreteness of the template bank: 
the search parameter space consists of cells of finite area centred on each template, but we only evaluate the
metric at the template itself.  For a given template, the $\Lambda_{\rm opt}$ value assigned to triggers should 
ideally arise from an integral over the area covered by the template.  In practice, artefacts due to finite cell 
size appear to be negligible except again near the $m_1=m_2$ line, where the re-weighting factor varies strongly 
from one template to the next, and for templates close to the large-$\Mc$ edge of the bank, mostly those with 
$m_1,m_2 >7.5\,\Msun$ (see Fig.~\ref{fig:lowmass_density}).  Such templates may be ranked anomalously highly 
considering that the areas they cover extend beyond the limits of the bank (at $\eta=0.25$ and $m_1+m_2=25\Msun$) 
where there are no simulated signals. 

Na\"\i vely applying the inverse metric formula (\ref{eq:Lambdaopt}) without regard to these caveats severely 
hurts the efficiency of our statistic, even if we ``remove'' the singularity by making sure no template has 
exactly equal masses; we find that statistic values in noise are dominated by a few triggers from the region
$m_1,m_2 >7.5\,\Msun$, causing the number of signals recovered at low false alarm rates to decline drastically.  
Clearly such templates, although they may be relatively likely to see signals, are not correctly described by our
first approximation. 

We choose to regularize the statistic values assigned to such templates by adding a small constant in quadrature 
to $\sqrt{\det\gamma(t_c,m_1,m_2)}$, such that the distribution of $\Lambda_{\rm opt}$ values in noise is not 
dominated by outliers from templates with small metric densities.  Having done so, we no longer see a decline in 
search efficiency at low false alarm rates.  We tested different choices of regularization constant and found 
that efficiency was optimised within a range of values for which $<1\%$ of noise triggers were significantly 
affected by regularization, and was insensitive to variation within this range.  
As remarked earlier, a more theoretically correct approach would be to model the variation of the prior over 
peaks of likelihood, including edge effects, in appropriate coordinates; however this would be significantly more 
complicated to implement and we see no evidence that doing so would lead to significant improvements in 
performance. 

\subsection{Improvement in search efficiency}

For each simulated signal, we calculate the false alarm rate (FAR) as the number out of the 5.2 million 
noise events assigned a ranking statistic (either matched filter $\rho$ or our optimized statistic 
$\Lambda_{\rm opt}$) equal to or higher than that of the signal trigger, divided by the search 
time.\footnote{Note that the resolution of the FAR value is limited by the search time; much smaller FARs 
can be assigned by multi-detector searches.} 

\begin{figure}
\includegraphics[width=0.5\textwidth]{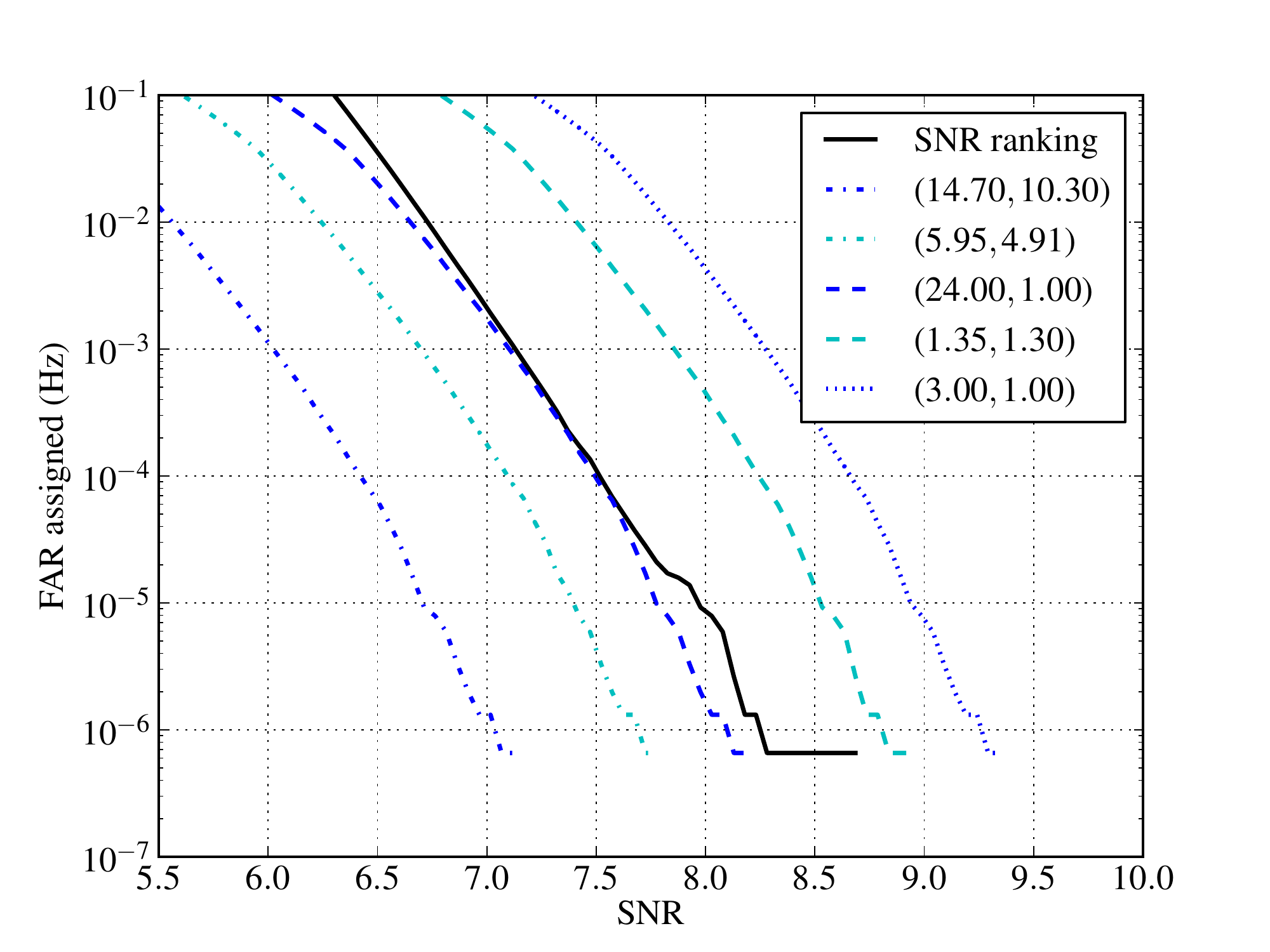}
\caption{False alarm rate (FAR) values assigned to signals with given matched filter SNR $\rho$. We show
FARs assigned using different ranking statistics, either $\rho$ (solid black line) or the likelihood ratio 
$\Lambda_{\rm opt}$ for a prior distribution of signals uniform over $(m_1,m_2)$ at constant SNR 
(dashed/dotted lines); for $\Lambda_{\rm opt}$ we show the FARs assigned for various possible signal 
component mass values, as in the legend.  
\label{fig:FAR_vs_SNR}
}
\end{figure}

We illustrate the relation between signal SNR and the resulting FAR value in Figure~\ref{fig:FAR_vs_SNR}, 
where for the likelihood ratio statistic we have chosen some representative points in the template bank to 
illustrate the effect of mass-dependent re-weighting. There is a non-trivial difference between the matched 
filter values for different candidate masses that result in the same FAR; thus, the search has a higher
sensitive volume (at fixed FAR) in regions of parameter space where the rate of signals relative to noise 
events is higher, and vice versa. Our mass-dependent statistic leads to an overall improvement in search 
efficiency since the gain in the expected number of signals from regions where the sensitivity is increased 
(relative to ranking by SNR) outweighs the loss from regions where it decreases.

\begin{figure}
\includegraphics[width=0.5\textwidth]{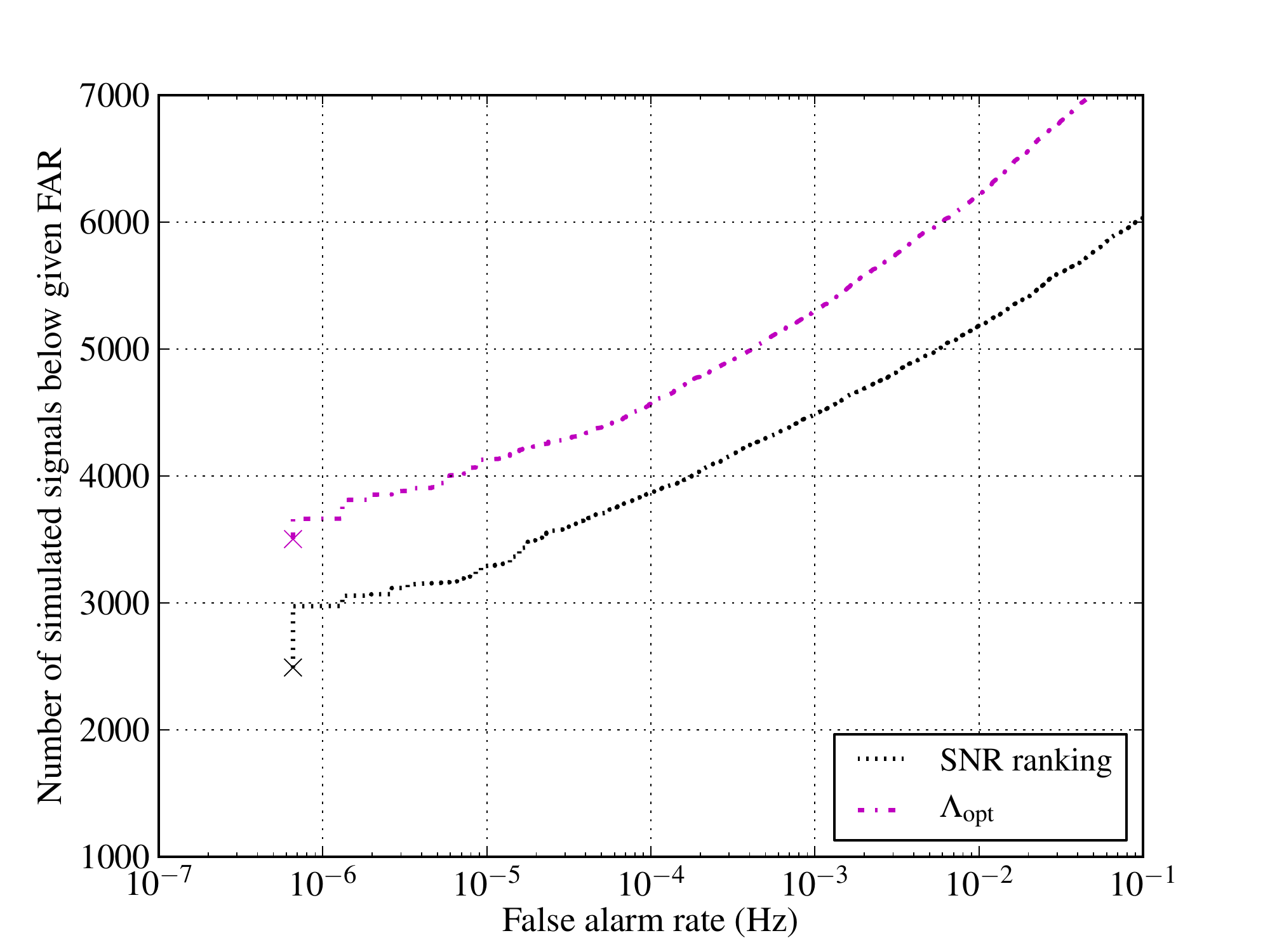}
\caption{Number of simulated signals recovered at a given false alarm rate (FAR) 
using either the matched filter SNR, or the optimized statistic $\Lambda_{\rm opt}$ constructed using the 
3-d metric $\gamma(t_c,m_1,m_2)$ for 2PN inspiral filters, as ranking statistic.
\label{fig:ROC_likelihood_SNR}
}
\end{figure}

The net effect of the re-weighting is thus to increase the number of simulated signals seen with a given FAR 
value or lower. 
We show this improvement in Figure~\ref{fig:ROC_likelihood_SNR}
where 
we see that the search efficiency, for the given parameter space and signal distribution, is improved by 
some tens of percent. Note that the evaluation of such ROCs is subject to fluctuations due to small number 
statistics at the extreme low FAR end. 
This result can also be interpreted in terms of a reduction in false alarm rate: by using the optimized 
statistic, a given number of signals is found at or below a FAR which is approximately a factor 10 lower 
than when using $\rho$ as ranking statistic. 

Interestingly, if instead of the 2PN metric density we use the density of a 2-d metric with constant 
components over $\tau_0,\tau_3$ space to evaluate $\Lambda_{\rm opt}$, we obtain almost the same ROC curve
with only a small loss of efficiency relative to the theoretically correct 2PN metric at low FAR. 
The great majority of the improvement over SNR ranking is thus due to the strong mass-dependence of the 
Jacobian of Eq.~(\ref{detJ_m1m2_tau0tau3}). 

We performed an independent check for the optimal classification of noise and simulated signal triggers 
by constructing a random forest classifier \cite{RandomForest,scikit-learn} with stratified 
cross-validation, where both sets of triggers are divided into 5 equally-sized ``folds'' and the probability 
of signal \emph{vs}.\ noise is evaluated for triggers in each fold using a forest trained on the remaining four, 
using the matched filter SNR and (simple functions of) the binary component masses as classification features. 
Using the forest predicted probability as ranking statistic, we found the efficiency as a function of FAR was 
within 2-3\% of that obtained by our $\Lambda_{\rm opt}$ statistic for FAR values above a 
few\,$\times10^{-6}$\,Hz; at lower FARs the random forest had somewhat worse efficiency, possibly due to the 
difficulty of training in regions of feature space with very few noise triggers.  This result gives us more 
confidence that the statistic we have implemented is close to optimal, at least within the statistical 
limitations of our example.  
Note that, even after imposing a cut on SNR 
to reduce the total trigger count, training a classifier with 200 trees required order(1/2 hour) computing 
time compared to a few minutes for evaluation of $\Lambda_{\rm opt}$ using pre-computed metric components for 
all 5.2 million noise triggers.

\subsection{Effect of varying signal distributions}

A natural question is how far the benefits produced by using an optimized statistic rather than the matched 
filter SNR are sensitive to a discrepancy between the true distribution of signals in nature and the 
distribution assumed in constructing the statistics.  Note too that the goal of this paper is to describe a 
\emph{method} of optimizing the ranking statistic of a search for \emph{any} given prior signal distribution; 
we make no claims that the prior uniform over $(m_1,m_2)$ at fixed SNR, which we chose for purposes of 
illustration, is in fact optimal for the true astrophysical distribution of binaries. 

We can address the issue of discrepancies in the assumed \emph{vs.}\ real signal distributions either by 
changing the prior $R(\mvec)D_{\rm hor}(\mvec)^3$ used in evaluating the $\Lambda_{\rm opt}$ statistic, but 
keeping the same set of simulated signals to evaluate efficiency; or by keeping the dependence of the ranking
statistic on $\mvec$ unchanged, but changing the weighting of simulated signals (or using a different set of 
simulations entirely) to correspond to a different true distribution.  Many different possible distributions of 
binary coalescence parameters have been proposed (see \cite{DominikEtAl:2012} for a recent study) and we will 
only consider a few simple scenarios here. 
\begin{figure}
\begin{center}
\hspace*{-0.5cm}
\includegraphics[width=0.51\textwidth]{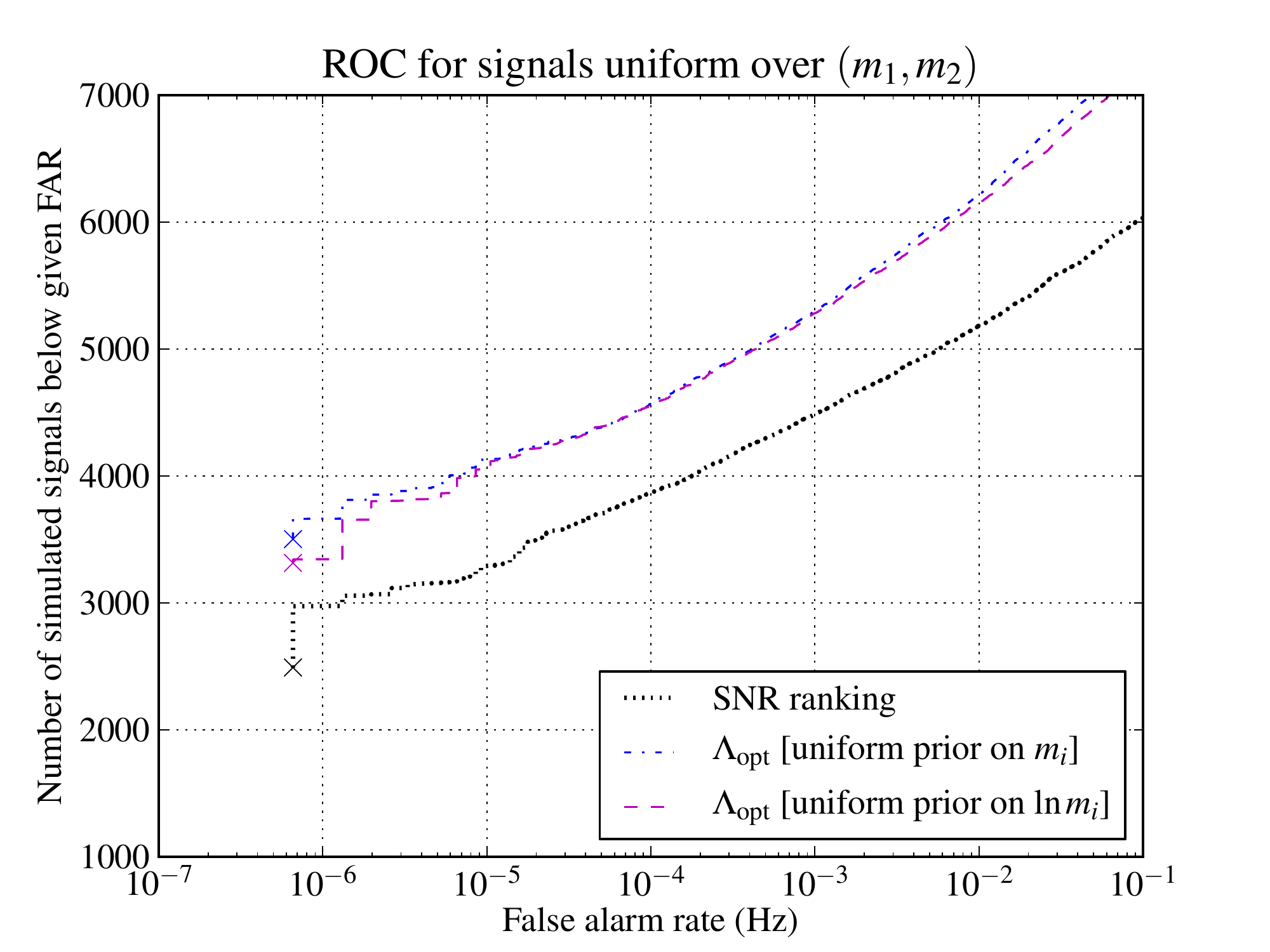} 
\includegraphics[width=0.51\textwidth]{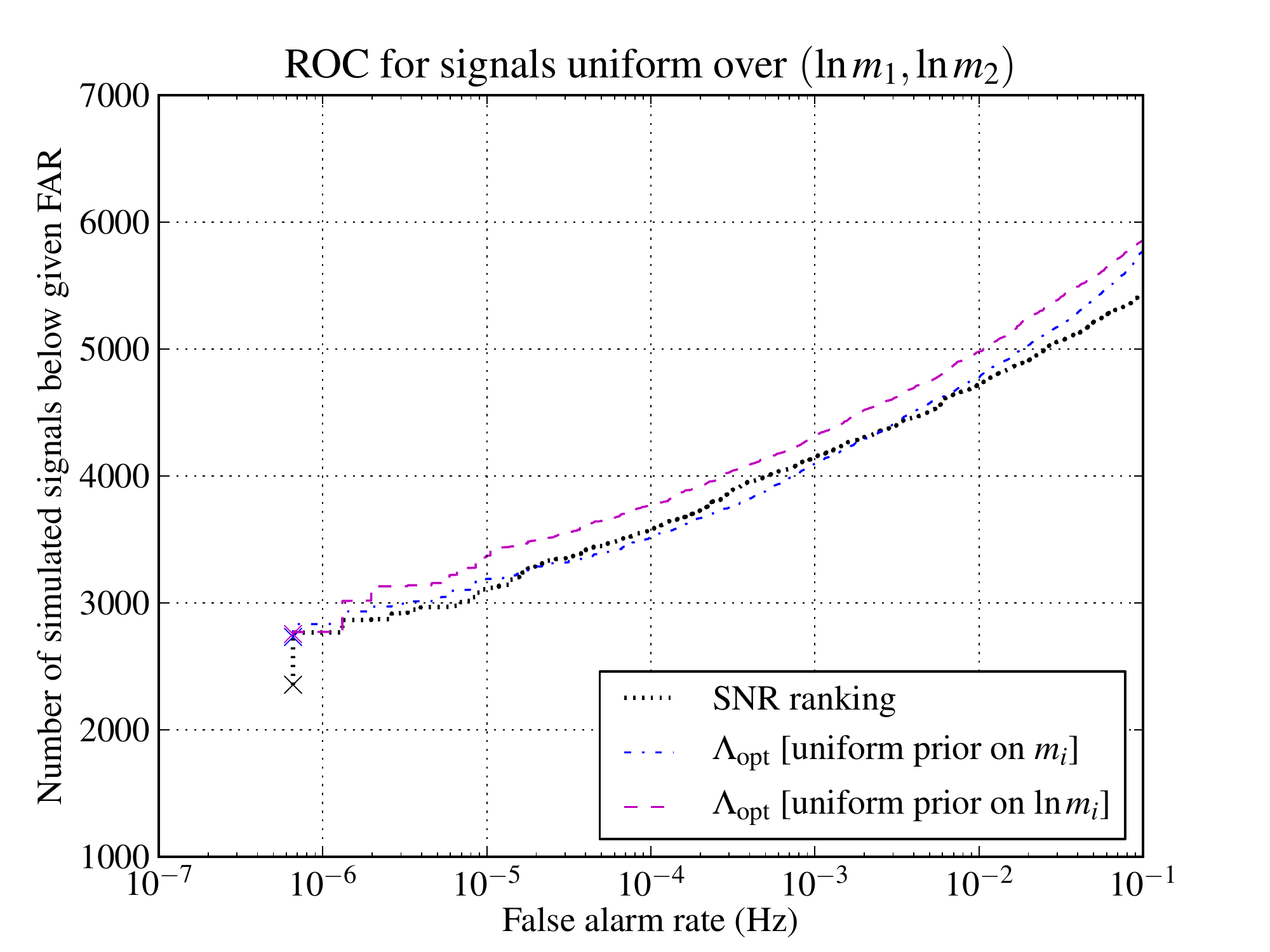} 
\hspace*{-0.5cm}
\end{center}
\vspace*{-0.6cm}
\caption{
\emph{Left}---Comparison of matched filter SNR and likelihood statistics constructed using different 
assumed prior signal distributions (either uniform in $(m_1,m_2)$ or uniform in $(\log m_1,\log m_2)$); 
ROCs are calculated with the injected signal distribution which is uniform over $m_1,m_2$. 
\emph{Right}---Comparison of statistics, as in the left panel: here, ROCs are
calculated by re-weighting the injected signal distribution to simulate a population uniform over 
$(\log m_1,\log m_2)$.
\vspace*{-0.2cm}
\label{fig:ROC_varydist}
}
\end{figure}

Figure~\ref{fig:ROC_varydist} shows the results of our comparisons. The left plot shows that, for a 
population of signals uniform over $(m_1,m_2)$, the statistic obtained by optimizing for a distribution 
uniform over $(\log m_1,\log m_2)$ shows almost the same gain in efficiency as the ``correctly'' optimized
statistic for the true signal population.\footnote{The optimization in either case is not perfect, due to 
the lower PN order used for the metric evaluation compared to the filters, and other approximations made in
deriving the statistic.} 
In the right plot we evaluate the ROC by totalling the simulated signals recovered with a FAR below threshold, 
with a parameter-dependent weight proportional to $1/(m_1m_2)$; the resulting figure of merit is appropriate 
to a true signal population uniform on $(\log m_1,\log m_2)$. We see in this case, as expected, the statistic
optimized using the ``correct'' prior, i.e.\ uniform over $(\log m_1,\log m_2)$, has the best efficiency; 
the statistic optimized for a signal distribution uniform over $(m_1,m_2)$ performs somewhat worse, but not
significantly worse than the matched filter SNR. The differences in efficiency are not large for this case;
note that the signal distribution uniform on $(\log m_1,\log m_2)$ is intermediate between the metric density 
(i.e.\ the prior distribution implied by ranking on SNR) and the uniform-over-component-mass distribution, 
in terms of how strongly lower-mass signals are favoured.

More extreme ``worse case'' scenarios can be conceived where the true signal population deviates further from 
the prior used to evaluate the $\Lambda_{\rm opt}$ statistic, and we would expect the efficiency of a search 
using such an ``unrealistic'' prior to suffer accordingly.  The limiting case would be for the entire 
population of real signals to be concentrated at the point where the assumed prior is smallest.  There would, 
nevertheless, be a nonzero lower bound on the efficiency of the search, provided that the prior did not vanish 
at that point. 

An example of a distribution which differs very strongly from an assumed prior uniform
over component masses is one where \emph{only binary neutron star} (BNS) coalescences occur, i.e.\ black hole 
binaries are entirely absent.\footnote{This example was suggested by the referee.}  
As shown in Figure~\ref{fig:FAR_vs_SNR}, the SNR required for a BNS of masses $(1.35,1.3)$ to be assigned a 
given false alarm rate below $\sim 10^{-5}$, using our statistic optimized for a uniform mass distribution, 
is greater than that required using SNR as ranking statistic, but only by approximately 0.5.  Thus in this case,
using a statistic optimised for a uniform mass distribution (including black hole binaries) will perform worse 
than the implied prior from the template metric, detecting a factor of approximately $~(8/8.5)^3\simeq 0.83$ 
fewer signals at low FAR.
Detailed discussions of the ``best'' choice of search prior given widely differing possible true astrophysical
distributions lie beyond the scope of this paper, but we make a few further remarks in the concluding Section. 

\section{Discussion} \label{sec:concl}
\noindent
We have shown that the optimal statistic for detection of transient gravitational wave signals of known form
over a broad intrinsic parameter space is \emph{not} the matched filter SNR, even in Gaussian noise, so long
as the distribution of signals over the parameter space is not identical to the metric density associated
with the matched filter. 
Since the signal distribution is determined by astrophysical processes, it is unlikely to be close to this 
default distribution; thus, a optimized Bayes factor detection statistic can in general outperform the 
matched filter SNR. We showed how to approximately evaluate this statistic in a search for coalescing 
binaries with non-spinning components, given only the information contained in ``triggers'' output by the 
standard LIGO inspiral search code. Extending the method to searches for binaries with non-precessing 
component spins is straightforward, given the recent development of metric-based methods \cite{Brown:2012qf} 
for covering the higher-dimensional space of intrinsic parameters required in this case. 

\paragraph*{Waveform uncertainty and parameter bias}
In the proof-of-principle analysis performed in this paper we have chosen to simulate signals using the 
TaylorT2 approximant and perform the matched filter search using a stationary-phase approximation akin to 
TaylorF2~\cite{DrozSPA:1999}, both at second post-Newtonian order.  This choice ensures that the parameter bias 
on the maximum likelihood mass template will be relatively small~\cite{Nitz:2013,BIOPS}.  
No choice of model template for a search will correspond precisely to the true gravitational waveform, so in 
general the distribution of recovered signal triggers in intrinsic parameter space will not correspond to the 
true distribution of sources.  In addition to reductions in SNR caused by an imperfect fitting factor between 
the true waveform and the filter, intrinsic parameter bias will reduce the efficiency of the search if it leads 
to an inaccurate assignment of ranking statistic values and if the statistic is optimized for the true signal 
distribution rather than the recovered one. However, the loss of efficiency caused by parameter bias is probably 
less important than the sub-optimality caused by our lack of knowledge of the true astrophysical distribution 
(see below). 

For the non-spinning template bank used in our example, bias in parameter recovery might not greatly affect the 
efficiency of our statistic:\footnote{We in fact started our case study using TaylorT4 simulated signal 
waveforms, which have a significant bias relative to the TaylorF2 templates; despite this we obtained a 
comparable gain in search efficiency.} the errors on recovered chirp mass $\Mc$ are expected to be very small, 
and it may be verified that the metric density (in component mass coordinates) varies weakly along lines of 
constant $\Mc$ 
except for a small region near the equal-mass line.

\paragraph*{Alternative marginalization methods}
An alternative path to evaluating the Bayes factor (marginalized likelihood ratio) would be to use the
template bank as a set of sampling points to explicitly evaluate the integral over intrinsic parameters 
$\mvec$ for short time intervals and restricted regions of parameter space. Thus, instead of taking the 
maximum of SNR in a trigger cluster to represent the entire peak of likelihood, one would add up the 
likelihood contributions over templates in a cluster and record the total, multiplying by the prior density 
(in appropriate coordinates) to obtain the analogue of Eq.~(\ref{eq:Lambdaopt}). Whether this method is 
preferable to approximating the integrand by a multivariate Gaussian will depend on technical issues such 
as the density of the bank, the accuracy to which it fulfils the minimal match criterion and the accuracy
to which the metric is known. For instance explicit integration over templates may be more suitable for a 
search with phenomenological inspiral-merger-ringdown filters where a metric cannot readily be calculated. 

\paragraph*{Multi-detector search} 
Our method leads to an optimized statistic when considering parameters affecting the signal seen in a single 
interferometric detector. The re-weighting procedure over the ``intrinsic'' parameters (masses and spins) 
remains essentially the same when considering more than one (spatially separated) detector. 
However, additional information is available from a multi-detector search, namely the differences in 
coalescence time and phase, and the relative signal amplitudes between detectors. The signal population will 
have a (possibly strongly) different distribution from that of noise events over these parameters; thus, 
knowledge of these distributions will help in distinguishing between them, and should further improve search 
efficiency relative to the standard multi-detector statistic (the quadrature sum of single-detector 
matched filter values \cite{ihope}).

Exploiting this information is a more complex task which we leave to future work. Machine learning 
classification appears the most practical strategy to deal with this multi-dimensional problem; in 
\cite{Cannon:2008zz}, a ``na\"\i ve Bayes'' framework was used to derive a ranking statistic using measured 
distributions over coincident event parameters for a modelled gravitational-wave burst search, although 
correlations between these parameters were not modelled. 

\paragraph*{Non-Gaussian noise} 
The presence of loud, unpredictable non-Gaussian artefacts (``glitches'') in the detector output 
\cite{Blackburn:2008ah} is a major challenge for attempts to calculate an optimal detection statistic from 
first principles. 
Various ad-hoc signal consistency tests have been implemented \cite{Allen:2004gu,ChadThesis} and shown to be 
effective in real noise at separating simulated signals from glitches, however they require tuning ``by hand'' 
to the data, and are not immediately connected to
the Bayesian framework described here. A theoretical approach to searching data containing outlying 
non-Gaussian samples was outlined in \cite{Allen:2002jw}, while the use of an explicit glitch model in 
evaluating Bayes factors was explored in \cite{Littenberg2010}; however so far an implementation suitable
to search extended periods of data for coalescing binary signals is not available. 

One approach using the existing trigger values calculated in the inspiral search algorithm, including the 
signal consistency tests, would be to simply treat these as additional parameters and attempt to estimate 
(fit) the relative signal and noise distributions empirically, along the lines of Section 
\ref{triggerstats}.\footnote{Ref.~\cite{Biswas:2012tv} describes an approach similar in spirit, but without 
explicitly modelling the trigger distributions.} 
The result would correspond to the Bayes factor calculated over some subset of the data corresponding to the 
quantities available. However, this might miss a considerable amount of information that could have helped 
exclude glitches; thus, the standard chi-squared test \cite{Allen:2004gu} does not distinguish between high- 
and low-frequency transients, although they may have very different distributions in any given stretch of 
data. 

No single method can be optimal for data containing arbitrary non-Gaussian artefacts, since the space of 
possible non-Gaussian behaviours is infinite; at most we can look for methods that approach optimality for 
populations of glitches resembling those seen in real data. 
Note that any such method must incorporate our findings on the density of Gaussian noise triggers, since
it must reduce to the optimal statistic described here in the Gaussian noise limit (for a single detector).

\paragraph*{Choice of prior and astrophysical uncertainties}
In the Introduction we mentioned the very large current uncertainties in the astrophysical distributions 
of coalescing binary systems. The question is then, what interpretation can we give to the prior distribution
over intrinsic parameters that enters our proposed optimal statistic. Two cases can be distinguished: 
pre-detection, where no significant additional information from GW observations is available;\footnote{Upper
limits from GW non-observations could \emph{somewhat} restrict the range of possible priors.} and 
post-detection, where the prior for future searches will be informed by the recovered parameter values
of some number of binary signals. 

In the pre-detection case one might choose the optimization prior to correspond to an astrophysical model
of compact binary formation and evolution (for instance \cite{DominikEtAl:2012}), after having marginalized 
over unknown nuisance parameters of the model. However, this approach has potential pitfalls, since one 
might mistakenly give a negligibly small weight to systems that the model (wrongly) predicts to be 
extremely rare, and thus produce a ``blind spot'' in the search; also, the complexity of such models is 
likely to make the search results difficult to interpret. Alternatively, one might want to verify the 
performance of an analysis pipeline before deployment by checking its performance in recovering simulated 
signals, \emph{i.e.}\ a Monte Carlo or mock data challenge, and demanding that the analysis be optimized for 
a well-defined signal population covering the intended parameter space \cite{Searle:2008}. 
As pointed out by \cite{Searle:2008ap}, for every analysis choice there may be \emph{some} signal 
distribution for which it is optimal, but this distribution may be either unphysical, implausible, or fail 
to correspond to the search requirements; use of an optimized statistic can avoid such issues. 

Post-detection, it will be possible to rule out some signal distributions, thus the range of possible search
priors will decrease. How to proceed in detail will depend on the scientific objectives: for instance one 
might aim to maximize the number of detected signals, or to obtain the best accuracy on measurements of 
astrophysical coalescence rates as a function of source parameters, or to obtain the largest Bayes factor in
distinguishing between different models of compact binary formation and evolution. We will leave such more 
complex optimization problems for future investigations. 

\section*{Acknowledgements} \noindent
We would like to thank Drew Keppel, Reinhard Prix, Andrew Lundgren, Ian Harry and Walter Del Pozzo for 
enlightening discussions; Will Farr for carefully reading an earlier draft of this paper;
the LIGO Scientific Collaboration for the generation and storage of the standard Gaussian data set used in this 
study; and the Albert-Einstein-Institut (Hannover) for the use of the Atlas computing cluster.
JV was supported by the research programme of the Foundation for Fundamental Research on Matter (FOM), which 
is partially supported by the Netherlands Organisation for Scientific Research (NWO); TD acknowledges
support from the Max-Planck-Gesellschaft. 

\appendix

\section{Marginalization over phase}
\label{app:coa}
\noindent
Here we will derive the phase-marginalized likelihood function and expand it around a maximum.
The full likelihood including phase is written in terms of the two phases of the signal $\hh_0$ and 
$\hh_{\pi/2}$:
\begin{align}
 L 
 &=\exp\left(-\frac{\A^2}{2}\Braket{\hh_0|\hh_0} + \A\Braket{\hh_0|d}\cos\coa + 
 \A\Braket{\hh_{\pi/2}|d}\sin\coa\right).
\end{align}
We marginalize over $\coa$ using a uniform prior $p(\coa|S)=(2\pi)^{-1}$ via the integral 
\begin{align}
 L' &= \exp\left(-\frac{\A^2}{2}\Braket{\hh|\hh}\right)
 \int_0^{2\pi}\frac{d\coa}{2\pi}\, \exp\left(\A\left[\Braket{\hh_0|d}\cos\coa + 
 \Braket{\hh_{\pi/2}|d}\sin\coa\right]\right),
\end{align}
which we perform by substituting $z=e^{i\coa}$ and $a=\A\Braket{\hh_0|d}+i\A\Braket{\hh_{\pi/2}|d}$ to arrive 
at a standard result in terms of the modified Bessel function $I_0$:
\begin{align}
 L'&=\frac{1}{2\pi}\exp\left(-\frac{\A^2}{2}\Braket{\hh|\hh}\right)\oint dz\exp\left(\frac{1}{2}(za+z^{-1}a^*) 
 \right)\\
 &=\exp\left[-\frac{\A^2}{2}\Braket{\hh|\hh}+\log I_0\left(\A\sqrt{\braket{\hh_0|d}^2+\braket{\hh_{\pi/2}|d}^2} 
 \right)\right].
\end{align}
\subsection{Maximum likelihood location}
In general, the parameters which maximize $L$ can be different from those that maximize $L'$, so we need to 
recalculate them to check that our use of the matched filter maximum is still valid. At the maximum 
we have
\begin{align}
 0 &=\partial_\alpha\left[-\frac{\A^2}{2}\Braket{\hh|\hh}\right] + \frac{I_1\left( \A\rho(\hh) \right)}
 {I_0\left( \A\rho(\hh) \right)} \partial_\alpha \left[\A\rho(\hh)\right],\label{eq:maxLcondition}
\end{align}
where $\rho(\hh) \equiv\sqrt{\braket{\hh_0|d}^2+\braket{\hh_{\pi/2}|d}^2}$ and $\alpha$ indexes all parameters
except $\coa$, including the amplitude.
First we consider the amplitude $\A$, where we find the maximum at $\A'_\ml$ given by the solution of 
\begin{align}
 \A_\ml' &= \frac{I_1\left( \A'_\ml\rho(\hh^\ml) \right)}{I_0\left( \A'_\ml\rho(\hh^\ml) \right)}
 \rho(\hh^\ml), 
\end{align}
such that 
\begin{align}
 \lim_{\A_\ml\to\infty}\A'_\ml&= \rho(\hh^\ml) = \A_\ml.
\end{align}
As the ratio of modified Bessel functions tends to $1$ for above-threshold detection statistic values (e.g.\ 
$I_1(36)/I_0(36)\approx 0.986$ for SNR=6), in this limit we recover the same maximum likelihood point as for 
the unmarginalized $L$ function. However, in the low SNR limit the recovered maximum $\A'_\ml<\A_\ml$.

For the intrinsic parameters labelled by $p$ we use Eq.~(\ref{eq:maxLcondition}) at the maximum likelihood 
point to find
\begin{align}
 \partial_p\log L'_\ml &= - \A_\ml^2 \Braket{\partial_p\hh_0|\hh_0} + 
 \frac{I_1\left(\A_\ml^2\right)}{I_0\left(\A_\ml^2\right)}
 \left[\Braket{\hh_0|d}\Braket{\partial_p\hh_0|d} + 
 \Braket{\hh_{\pi/2}|d}\Braket{\partial_p\hh_{\pi/2}|d}\right].
\end{align}
Using the normalisation condition which implies $\Braket{\partial_p\hh|\hh}=0$ for all parameter values, we 
also find
\begin{gather} \label{eq:d2hh}
 0 =\partial_p\partial_q\Braket{\hh|\hh} = \Braket{\partial_p\partial_q \hh|\hh}+
 \Braket{\partial_p \hh|\partial_q \hh}, \\
 \Braket{\hh_0|d}\Braket{\partial_p\hh_0|d}_\ml = -\Braket{\hh_{\pi/2}|d}\Braket{\partial_p\hh_{\pi/2}|d}_\ml
 \qquad \implies \qquad \partial_p\, \rho(\hh)_{|\ml} = 0. \label{eq:dLdp}
\end{gather}

\subsection{Expansion around maximum} \noindent
In order to expand $\log L'$ we also need the second derivatives with respect to the parameter values.
For the amplitude term $\partial_\A\partial_\A$ we have
\begin{align}
\partial_\A\partial_\A \log L' &\approx -1 + \A_\ml^2 
 \left(1+\frac{I_2(\A_\ml^2) }{I_0(\A_\ml^2)} - 2\frac{I_1(\A_\ml^2)^2}{I_0(\A_\ml^2)^2}\right),
\end{align}
where the second term tends to zero at high SNR, so $\partial_\A \partial_\A \log L' \approx -1$.
For the intrinsic parameters we obtain
\begin{multline}
 \partial_p\partial_q\log L_\ml' = \frac{I_1(\A_\ml^2)}{I_0(\A_\ml^2)}\: \cdot \\ 
 \left[ \Braket{\hh_0|d}\Braket{\partial_p\partial_q\hh_0|d} + 
 \Braket{\hh_{\pi/2}|d}\Braket{\partial_p\partial_q\hh_{\pi/2}|d} + 
 \Braket{\partial_p\hh_0|d}\Braket{\partial_q\hh_0|d} + 
 \Braket{\partial_p\hh_{\pi/2}|d}\Braket{\partial_q\hh_{\pi/2}|d} \right]_\ml,  
\end{multline}
where derivatives of the Bessel functions vanish at the ML point via (\ref{eq:dLdp}), which also implies
$\partial_\A \partial_p \log L' = 0$ at this point.
To proceed we write
\begin{align}
 d = \A_\ml \hh^\ml + n = \A_\ml (\cos\phi_c^\ml \hh_0 +\sin\phi_c^\ml \hh_{\pi/2} ) + n
\end{align}
where $\tan\phi_c^\ml=\braket{\hh^\ml_{\pi/2}|d}/\braket{\hh^\ml_0|d}$ and $n$ is a noise residual having 
vanishing overlap with $\hh^\ml$, which is formally of sub-leading order in $\A_\ml$ and can be neglected for a
sufficiently high SNR \cite{PoissonWill}.\footnote{Note that this may not hold in
the presence of a signal which deviates significantly from the model $\hh(\pvec)$.} 
We then find, after some rearrangement and applying (\ref{eq:dLdp}),
\begin{align} \label{eq:dpdq_logL_ML}
 \partial_p\partial_q\log L_\ml' &\simeq \frac{I_1(\A_\ml^2)}{I_0(\A_\ml^2)}
 \left[ -\A_\ml^2 \Braket{\partial_p\hh_0|\partial_q\hh_0} + \frac{\braket{\partial_p\hh_0|d}
 \braket{\partial_q\hh_0|d}}{\sin^2\phi_c^\ml} \right]_\ml  \nonumber \\
 &\simeq -\A_\ml^2 \frac{I_1(\A_\ml^2)}{I_0(\A_\ml^2)} \gamma_{pq}(\hh^\ml), 
\end{align}
where $\gamma_{pq} = \braket{\partial_p\hh_0|\partial_q\hh_0}-\braket{\partial_p\hh_0|\hh_{\pi/2}}
\braket{\partial_q\hh_0|\hh_{\pi/2}}$ is the \emph{phase-projected metric} describing overlaps maximized
over phase difference. 
We then find the expansion of the likelihood 
\begin{align}
 \log L'(\A_\ml+\Delta\A,\pvec^\ml+\Delta\pvec) &\simeq -\frac{\A_\ml^2}{2} - \frac{\Delta\A^2}{2} 
 + \log I_0\left(\A_\ml^2\right) + \frac{1}{2}\Delta\theta_p\Delta\theta_q 
 \left(\partial_p\partial_q \log L'\right)_\ml.
\end{align}
With the assumption that the priors are slowly varying around the maximum, we obtain, for an n-dimensional
parameter vector $\pvec$,  
\begin{align}
 \int d\A\, d^n\pvec\, p(\A,\pvec|S) L' &\simeq p(\A,\pvec|S)_{\A_\ml,\pvec^\ml} L_\ml' 
 \int d(\Delta \A)\, e^{-\Delta\A^2/2} \int d^n\pvec\, \exp\left[ -\frac{1}{2}\A_\ml^2\gamma'_{pq}
 \Delta\theta_p\Delta\theta_q \right] \\
 &= p(\A_\ml,\pvec^\ml|S) \exp\left( -\frac{\A_\ml^2}{2}\right) I_0\left(\A_\ml^2\right) 
 \A_\ml^{-n} \sqrt{\frac{(2\pi)^{n+1}}{\det\gamma'(\pvec^\ml)}} ,
\end{align}
where $\gamma'$ is equal to $\gamma$ rescaled by the ratio of Bessel functions appearing in 
Eq.~(\ref{eq:dpdq_logL_ML}); for moderate or large SNRs this ratio is very close to unity, hence we may 
approximate $\gamma'\simeq\gamma$.

\bibliography{biblio}

\begin{thebibliography}{49}%
\makeatletter
\providecommand \@ifxundefined [1]{%
 \@ifx{#1\undefined}
}%
\providecommand \@ifnum [1]{%
 \ifnum #1\expandafter \@firstoftwo
 \else \expandafter \@secondoftwo
 \fi
}%
\providecommand \@ifx [1]{%
 \ifx #1\expandafter \@firstoftwo
 \else \expandafter \@secondoftwo
 \fi
}%
\providecommand \natexlab [1]{#1}%
\providecommand \enquote  [1]{``#1''}%
\providecommand \bibnamefont  [1]{#1}%
\providecommand \bibfnamefont [1]{#1}%
\providecommand \citenamefont [1]{#1}%
\providecommand \href@noop [0]{\@secondoftwo}%
\providecommand \href [0]{\begingroup \@sanitize@url \@href}%
\providecommand \@href[1]{\@@startlink{#1}\@@href}%
\providecommand \@@href[1]{\endgroup#1\@@endlink}%
\providecommand \@sanitize@url [0]{\catcode `\\12\catcode `\$12\catcode
  `\&12\catcode `\#12\catcode `\^12\catcode `\_12\catcode `\%12\relax}%
\providecommand \@@startlink[1]{}%
\providecommand \@@endlink[0]{}%
\providecommand \url  [0]{\begingroup\@sanitize@url \@url }%
\providecommand \@url [1]{\endgroup\@href {#1}{\urlprefix }}%
\providecommand \urlprefix  [0]{URL }%
\providecommand \Eprint [0]{\href }%
\providecommand \doibase [0]{http://dx.doi.org/}%
\providecommand \selectlanguage [0]{\@gobble}%
\providecommand \bibinfo  [0]{\@secondoftwo}%
\providecommand \bibfield  [0]{\@secondoftwo}%
\providecommand \translation [1]{[#1]}%
\providecommand \BibitemOpen [0]{}%
\providecommand \bibitemStop [0]{}%
\providecommand \bibitemNoStop [0]{.\EOS\space}%
\providecommand \EOS [0]{\spacefactor3000\relax}%
\providecommand \BibitemShut  [1]{\csname bibitem#1\endcsname}%
\let\auto@bib@innerbib\@empty
\bibitem [{\citenamefont {Abadie}\ \emph {et~al.}(2010)\citenamefont {Abadie}
  \emph {et~al.}}]{LSCRates}%
  \BibitemOpen
  \bibfield  {author} {\bibinfo {author} {\bibfnamefont {J.}~\bibnamefont
  {Abadie}} \emph {et~al.} (\bibinfo {collaboration} {LIGO Scientific
  Collaboration, Virgo Collaboration}),\ }\href {\doibase
  10.1088/0264-9381/27/17/173001} {\bibfield  {journal} {\bibinfo  {journal}
  {Class. Quant. Grav.}\ }\textbf {\bibinfo {volume} {27}},\ \bibinfo {pages}
  {173001} (\bibinfo {year} {2010})},\ \Eprint {http://arxiv.org/abs/1003.2480}
  {arXiv:1003.2480} \BibitemShut {NoStop}%
\bibitem [{\citenamefont {Babak}\ \emph {et~al.}(2006)\citenamefont {Babak},
  \citenamefont {Balasubramanian}, \citenamefont {Churches}, \citenamefont
  {Cokelaer},\ and\ \citenamefont {Sathyaprakash}}]{Babak:2006ty}%
  \BibitemOpen
  \bibfield  {author} {\bibinfo {author} {\bibfnamefont {S.}~\bibnamefont
  {Babak}}, \bibinfo {author} {\bibfnamefont {R.}~\bibnamefont
  {Balasubramanian}}, \bibinfo {author} {\bibfnamefont {D.}~\bibnamefont
  {Churches}}, \bibinfo {author} {\bibfnamefont {T.}~\bibnamefont {Cokelaer}},
  \ and\ \bibinfo {author} {\bibfnamefont {B.~S.}\ \bibnamefont
  {Sathyaprakash}},\ }\href {\doibase 10.1088/0264-9381/23/18/002} {\bibfield
  {journal} {\bibinfo  {journal} {Class. Quant. Grav.}\ }\textbf {\bibinfo
  {volume} {23}},\ \bibinfo {pages} {5477} (\bibinfo {year} {2006})},\ \Eprint
  {http://arxiv.org/abs/gr-qc/0604037} {arXiv:gr-qc/0604037} \BibitemShut
  {NoStop}%
\bibitem [{\citenamefont {{Cokelaer}}(2007)}]{Cokelaer:2007}%
  \BibitemOpen
  \bibfield  {author} {\bibinfo {author} {\bibfnamefont {T.}~\bibnamefont
  {{Cokelaer}}},\ }\href {\doibase 10.1103/PhysRevD.76.102004} {\bibfield
  {journal} {\bibinfo  {journal} {\prd}\ }\textbf {\bibinfo {volume} {76}},\
  \bibinfo {eid} {102004} (\bibinfo {year} {2007})},\ \Eprint
  {http://arxiv.org/abs/0706.4437} {arXiv:0706.4437} \BibitemShut {NoStop}%
\bibitem [{\citenamefont {Keppel}\ \emph {et~al.}(2013)\citenamefont {Keppel},
  \citenamefont {Lundgren}, \citenamefont {Owen},\ and\ \citenamefont
  {Zhu}}]{Keppel:2013kia}%
  \BibitemOpen
  \bibfield  {author} {\bibinfo {author} {\bibfnamefont {D.}~\bibnamefont
  {Keppel}}, \bibinfo {author} {\bibfnamefont {A.~P.}\ \bibnamefont
  {Lundgren}}, \bibinfo {author} {\bibfnamefont {B.~J.}\ \bibnamefont {Owen}},
  \ and\ \bibinfo {author} {\bibfnamefont {H.}~\bibnamefont {Zhu}},\ }\href
  {\doibase 10.1103/PhysRevD.88.063002} {\bibfield  {journal} {\bibinfo
  {journal} {Phys.Rev.}\ }\textbf {\bibinfo {volume} {D88}},\ \bibinfo {pages}
  {063002} (\bibinfo {year} {2013})},\ \Eprint {http://arxiv.org/abs/1305.5381}
  {arXiv:1305.5381} \BibitemShut {NoStop}%
\bibitem [{\citenamefont {Brown}\ \emph
  {et~al.}(2012{\natexlab{a}})\citenamefont {Brown}, \citenamefont {Lundgren},\
  and\ \citenamefont {O'Shaughnessy}}]{Brown:2012gs}%
  \BibitemOpen
  \bibfield  {author} {\bibinfo {author} {\bibfnamefont {D.~A.}\ \bibnamefont
  {Brown}}, \bibinfo {author} {\bibfnamefont {A.}~\bibnamefont {Lundgren}}, \
  and\ \bibinfo {author} {\bibfnamefont {R.}~\bibnamefont {O'Shaughnessy}},\
  }\href {\doibase 10.1103/PhysRevD.86.064020} {\bibfield  {journal} {\bibinfo
  {journal} {Phys. Rev.}\ }\textbf {\bibinfo {volume} {D86}},\ \bibinfo {pages}
  {064020} (\bibinfo {year} {2012}{\natexlab{a}})},\ \Eprint
  {http://arxiv.org/abs/1203.6060} {arXiv:1203.6060} \BibitemShut {NoStop}%
\bibitem [{\citenamefont {Brown}\ \emph
  {et~al.}(2012{\natexlab{b}})\citenamefont {Brown}, \citenamefont {Harry},
  \citenamefont {Lundgren},\ and\ \citenamefont {Nitz}}]{Brown:2012qf}%
  \BibitemOpen
  \bibfield  {author} {\bibinfo {author} {\bibfnamefont {D.~A.}\ \bibnamefont
  {Brown}}, \bibinfo {author} {\bibfnamefont {I.}~\bibnamefont {Harry}},
  \bibinfo {author} {\bibfnamefont {A.}~\bibnamefont {Lundgren}}, \ and\
  \bibinfo {author} {\bibfnamefont {A.~H.}\ \bibnamefont {Nitz}},\ }\href
  {\doibase 10.1103/PhysRevD.86.084017} {\bibfield  {journal} {\bibinfo
  {journal} {Phys. Rev.}\ }\textbf {\bibinfo {volume} {D86}},\ \bibinfo {pages}
  {084017} (\bibinfo {year} {2012}{\natexlab{b}})},\ \Eprint
  {http://arxiv.org/abs/1207.6406} {arXiv:1207.6406} \BibitemShut {NoStop}%
\bibitem [{\citenamefont {Ajith}\ \emph {et~al.}(2012)\citenamefont {Ajith},
  \citenamefont {Fotopoulos}, \citenamefont {Privitera}, \citenamefont
  {Neunzert},\ and\ \citenamefont {Weinstein}}]{Ajith:2012mn}%
  \BibitemOpen
  \bibfield  {author} {\bibinfo {author} {\bibfnamefont {P.}~\bibnamefont
  {Ajith}}, \bibinfo {author} {\bibfnamefont {N.}~\bibnamefont {Fotopoulos}},
  \bibinfo {author} {\bibfnamefont {S.}~\bibnamefont {Privitera}}, \bibinfo
  {author} {\bibfnamefont {A.}~\bibnamefont {Neunzert}}, \ and\ \bibinfo
  {author} {\bibfnamefont {A.}~\bibnamefont {Weinstein}},\ }\href@noop {} {\
  (\bibinfo {year} {2012})},\ \Eprint {http://arxiv.org/abs/1210.6666}
  {arXiv:1210.6666} \BibitemShut {NoStop}%
\bibitem [{\citenamefont {Harry}\ \emph {et~al.}(2013)\citenamefont {Harry},
  \citenamefont {Nitz}, \citenamefont {Brown}, \citenamefont {Lundgren},
  \citenamefont {Ochsner} \emph {et~al.}}]{Harry:2013tca}%
  \BibitemOpen
  \bibfield  {author} {\bibinfo {author} {\bibfnamefont {I.}~\bibnamefont
  {Harry}}, \bibinfo {author} {\bibfnamefont {A.}~\bibnamefont {Nitz}},
  \bibinfo {author} {\bibfnamefont {D.~A.}\ \bibnamefont {Brown}}, \bibinfo
  {author} {\bibfnamefont {A.}~\bibnamefont {Lundgren}}, \bibinfo {author}
  {\bibfnamefont {E.}~\bibnamefont {Ochsner}},  \emph {et~al.},\ }\href@noop {}
  {\  (\bibinfo {year} {2013})},\ \Eprint {http://arxiv.org/abs/1307.3562}
  {arXiv:1307.3562} \BibitemShut {NoStop}%
\bibitem [{\citenamefont {Privitera}\ \emph {et~al.}(2014)\citenamefont
  {Privitera}, \citenamefont {Mohapatra}, \citenamefont {Ajith}, \citenamefont
  {Cannon}, \citenamefont {Fotopoulos} \emph {et~al.}}]{Privitera:2013xza}%
  \BibitemOpen
  \bibfield  {author} {\bibinfo {author} {\bibfnamefont {S.}~\bibnamefont
  {Privitera}}, \bibinfo {author} {\bibfnamefont {S.~R.~P.}\ \bibnamefont
  {Mohapatra}}, \bibinfo {author} {\bibfnamefont {P.}~\bibnamefont {Ajith}},
  \bibinfo {author} {\bibfnamefont {K.}~\bibnamefont {Cannon}}, \bibinfo
  {author} {\bibfnamefont {N.}~\bibnamefont {Fotopoulos}},  \emph {et~al.},\
  }\href {\doibase 10.1103/PhysRevD.89.024003} {\bibfield  {journal} {\bibinfo
  {journal} {Phys.Rev.}\ }\textbf {\bibinfo {volume} {D89}},\ \bibinfo {pages}
  {024003} (\bibinfo {year} {2014})},\ \Eprint {http://arxiv.org/abs/1310.5633}
  {arXiv:1310.5633 [gr-qc]} \BibitemShut {NoStop}%
\bibitem [{\citenamefont {Brown}(2004)}]{DuncanThesis}%
  \BibitemOpen
  \bibfield  {author} {\bibinfo {author} {\bibfnamefont {D.~A.}\ \bibnamefont
  {Brown}},\ }\emph {\bibinfo {title} {Search for gravitational radiation from
  black hole {MACHOs} in the {G}alactic halo}},\ \href@noop {} {Ph.D. thesis},\
  \bibinfo  {school} {University of Wisconsin--Milwaukee} (\bibinfo {year}
  {2004}),\ \Eprint {http://arxiv.org/abs/0705.1514} {arXiv:0705.1514 [gr-qc]}
  \BibitemShut {NoStop}%
\bibitem [{\citenamefont {Allen}\ \emph {et~al.}(2012)\citenamefont {Allen},
  \citenamefont {Anderson}, \citenamefont {Brady}, \citenamefont {Brown},\ and\
  \citenamefont {Creighton}}]{FindChirp}%
  \BibitemOpen
  \bibfield  {author} {\bibinfo {author} {\bibfnamefont {B.}~\bibnamefont
  {Allen}}, \bibinfo {author} {\bibfnamefont {W.~G.}\ \bibnamefont {Anderson}},
  \bibinfo {author} {\bibfnamefont {P.~R.}\ \bibnamefont {Brady}}, \bibinfo
  {author} {\bibfnamefont {D.~A.}\ \bibnamefont {Brown}}, \ and\ \bibinfo
  {author} {\bibfnamefont {J.~D.~E.}\ \bibnamefont {Creighton}},\ }\href
  {\doibase 10.1103/PhysRevD.85.122006} {\bibfield  {journal} {\bibinfo
  {journal} {Phys. Rev. D}\ }\textbf {\bibinfo {volume} {85}},\ \bibinfo
  {pages} {122006} (\bibinfo {year} {2012})}\BibitemShut {NoStop}%
\bibitem [{\citenamefont {Babak}\ \emph {et~al.}(2013)\citenamefont {Babak}
  \emph {et~al.}}]{ihope}%
  \BibitemOpen
  \bibfield  {author} {\bibinfo {author} {\bibfnamefont {S.}~\bibnamefont
  {Babak}} \emph {et~al.},\ }\href {\doibase 10.1103/PhysRevD.87.024033}
  {\bibfield  {journal} {\bibinfo  {journal} {Phys. Rev. D}\ }\textbf {\bibinfo
  {volume} {87}},\ \bibinfo {pages} {024033} (\bibinfo {year}
  {2013})}\BibitemShut {NoStop}%
\bibitem [{\citenamefont {Abbott}\ \emph {et~al.}(2009)\citenamefont {Abbott}
  \emph {et~al.}}]{S5Year1}%
  \BibitemOpen
  \bibfield  {author} {\bibinfo {author} {\bibfnamefont {B.}~\bibnamefont
  {Abbott}} \emph {et~al.} (\bibinfo {collaboration} {LIGO Scientific
  Collaboration}),\ }\href {\doibase 10.1103/PhysRevD.79.122001} {\bibfield
  {journal} {\bibinfo  {journal} {Phys. Rev.}\ }\textbf {\bibinfo {volume}
  {D79}},\ \bibinfo {pages} {122001} (\bibinfo {year} {2009})},\ \Eprint
  {http://arxiv.org/abs/0901.0302} {arXiv:0901.0302} \BibitemShut {NoStop}%
\bibitem [{\citenamefont {Balasubramanian}\ \emph {et~al.}(1996)\citenamefont
  {Balasubramanian}, \citenamefont {Sathyaprakash},\ and\ \citenamefont
  {Dhurandhar}}]{Balasubramanian:1995bm}%
  \BibitemOpen
  \bibfield  {author} {\bibinfo {author} {\bibfnamefont {R.}~\bibnamefont
  {Balasubramanian}}, \bibinfo {author} {\bibfnamefont {B.}~\bibnamefont
  {Sathyaprakash}}, \ and\ \bibinfo {author} {\bibfnamefont {S.}~\bibnamefont
  {Dhurandhar}},\ }\href {\doibase 10.1103/PhysRevD.54.1860.2,
  10.1103/PhysRevD.53.3033} {\bibfield  {journal} {\bibinfo  {journal} {Phys.
  Rev.}\ }\textbf {\bibinfo {volume} {D53}},\ \bibinfo {pages} {3033} (\bibinfo
  {year} {1996})},\ \Eprint {http://arxiv.org/abs/gr-qc/9508011}
  {arXiv:gr-qc/9508011} \BibitemShut {NoStop}%
\bibitem [{\citenamefont {Owen}(1996)}]{Owen:1996}%
  \BibitemOpen
  \bibfield  {author} {\bibinfo {author} {\bibfnamefont {B.~J.}\ \bibnamefont
  {Owen}},\ }\href {\doibase 10.1103/PhysRevD.53.6749} {\bibfield  {journal}
  {\bibinfo  {journal} {Phys. Rev.}\ }\textbf {\bibinfo {volume} {D53}},\
  \bibinfo {pages} {6749} (\bibinfo {year} {1996})},\ \Eprint
  {http://arxiv.org/abs/gr-qc/9511032} {arXiv:gr-qc/9511032} \BibitemShut
  {NoStop}%
\bibitem [{\citenamefont {Owen}\ and\ \citenamefont
  {Sathyaprakash}(1999)}]{OwenSathya:1999}%
  \BibitemOpen
  \bibfield  {author} {\bibinfo {author} {\bibfnamefont {B.~J.}\ \bibnamefont
  {Owen}}\ and\ \bibinfo {author} {\bibfnamefont {B.~S.}\ \bibnamefont
  {Sathyaprakash}},\ }\href {\doibase 10.1103/PhysRevD.60.022002} {\bibfield
  {journal} {\bibinfo  {journal} {Phys. Rev. D}\ }\textbf {\bibinfo {volume}
  {60}},\ \bibinfo {pages} {022002} (\bibinfo {year} {1999})}\BibitemShut
  {NoStop}%
\bibitem [{\citenamefont {Finn}\ and\ \citenamefont
  {Chernoff}(1993)}]{FinnChernoff:1993}%
  \BibitemOpen
  \bibfield  {author} {\bibinfo {author} {\bibfnamefont {L.~S.}\ \bibnamefont
  {Finn}}\ and\ \bibinfo {author} {\bibfnamefont {D.~F.}\ \bibnamefont
  {Chernoff}},\ }\href {\doibase 10.1103/PhysRevD.47.2198} {\bibfield
  {journal} {\bibinfo  {journal} {Phys. Rev. D}\ }\textbf {\bibinfo {volume}
  {47}},\ \bibinfo {pages} {2198} (\bibinfo {year} {1993})}\BibitemShut
  {NoStop}%
\bibitem [{\citenamefont {Searle}\ \emph {et~al.}(2009)\citenamefont {Searle},
  \citenamefont {Sutton},\ and\ \citenamefont {Tinto}}]{Searle:2008ap}%
  \BibitemOpen
  \bibfield  {author} {\bibinfo {author} {\bibfnamefont {A.~C.}\ \bibnamefont
  {Searle}}, \bibinfo {author} {\bibfnamefont {P.~J.}\ \bibnamefont {Sutton}},
  \ and\ \bibinfo {author} {\bibfnamefont {M.}~\bibnamefont {Tinto}},\ }\href
  {\doibase 10.1088/0264-9381/26/15/155017} {\bibfield  {journal} {\bibinfo
  {journal} {Class. Quant. Grav.}\ }\textbf {\bibinfo {volume} {26}},\ \bibinfo
  {pages} {155017} (\bibinfo {year} {2009})},\ \Eprint
  {http://arxiv.org/abs/0809.2809} {arXiv:0809.2809} \BibitemShut {NoStop}%
\bibitem [{\citenamefont {{Prix}}\ and\ \citenamefont
  {{Krishnan}}(2009)}]{Prix:2009}%
  \BibitemOpen
  \bibfield  {author} {\bibinfo {author} {\bibfnamefont {R.}~\bibnamefont
  {{Prix}}}\ and\ \bibinfo {author} {\bibfnamefont {B.}~\bibnamefont
  {{Krishnan}}},\ }\href {\doibase 10.1088/0264-9381/26/20/204013} {\bibfield
  {journal} {\bibinfo  {journal} {Class. Quant. Grav.}\ }\textbf {\bibinfo
  {volume} {26}},\ \bibinfo {eid} {204013} (\bibinfo {year} {2009})},\ \Eprint
  {http://arxiv.org/abs/0907.2569} {arXiv:0907.2569} \BibitemShut {NoStop}%
\bibitem [{\citenamefont {Biswas}\ \emph {et~al.}(2012)\citenamefont {Biswas},
  \citenamefont {Brady}, \citenamefont {Burguet-Castell}, \citenamefont
  {Cannon}, \citenamefont {Clayton} \emph {et~al.}}]{Biswas:2012tv}%
  \BibitemOpen
  \bibfield  {author} {\bibinfo {author} {\bibfnamefont {R.}~\bibnamefont
  {Biswas}}, \bibinfo {author} {\bibfnamefont {P.~R.}\ \bibnamefont {Brady}},
  \bibinfo {author} {\bibfnamefont {J.}~\bibnamefont {Burguet-Castell}},
  \bibinfo {author} {\bibfnamefont {K.}~\bibnamefont {Cannon}}, \bibinfo
  {author} {\bibfnamefont {J.}~\bibnamefont {Clayton}},  \emph {et~al.},\
  }\href {\doibase 10.1103/PhysRevD.85.122008} {\bibfield  {journal} {\bibinfo
  {journal} {Phys. Rev.}\ }\textbf {\bibinfo {volume} {D85}},\ \bibinfo {pages}
  {122008} (\bibinfo {year} {2012})},\ \Eprint {http://arxiv.org/abs/1201.2959}
  {arXiv:1201.2959} \BibitemShut {NoStop}%
\bibitem [{\citenamefont {Cannon}\ \emph {et~al.}(2013)\citenamefont {Cannon},
  \citenamefont {Hanna},\ and\ \citenamefont {Keppel}}]{Cannon:2012zt}%
  \BibitemOpen
  \bibfield  {author} {\bibinfo {author} {\bibfnamefont {K.}~\bibnamefont
  {Cannon}}, \bibinfo {author} {\bibfnamefont {C.}~\bibnamefont {Hanna}}, \
  and\ \bibinfo {author} {\bibfnamefont {D.}~\bibnamefont {Keppel}},\ }\href
  {\doibase 10.1103/PhysRevD.88.024025} {\bibfield  {journal} {\bibinfo
  {journal} {Phys. Rev.}\ }\textbf {\bibinfo {volume} {D88}},\ \bibinfo {pages}
  {024025} (\bibinfo {year} {2013})},\ \Eprint {http://arxiv.org/abs/1209.0718}
  {arXiv:1209.0718} \BibitemShut {NoStop}%
\bibitem [{\citenamefont {{Finn}}(1992)}]{Finn:1992}%
  \BibitemOpen
  \bibfield  {author} {\bibinfo {author} {\bibfnamefont {L.~S.}\ \bibnamefont
  {{Finn}}},\ }\href {\doibase 10.1103/PhysRevD.46.5236} {\bibfield  {journal}
  {\bibinfo  {journal} {\prd}\ }\textbf {\bibinfo {volume} {46}},\ \bibinfo
  {pages} {5236} (\bibinfo {year} {1992})},\ \Eprint
  {http://arxiv.org/abs/arXiv:gr-qc/9209010} {arXiv:gr-qc/9209010} \BibitemShut
  {NoStop}%
\bibitem [{\citenamefont {Cutler}\ and\ \citenamefont
  {Flanagan}(1994)}]{CutlerFlanagan:1994}%
  \BibitemOpen
  \bibfield  {author} {\bibinfo {author} {\bibfnamefont {C.}~\bibnamefont
  {Cutler}}\ and\ \bibinfo {author} {\bibfnamefont {E.~E.}\ \bibnamefont
  {Flanagan}},\ }\href {\doibase 10.1103/PhysRevD.49.2658} {\bibfield
  {journal} {\bibinfo  {journal} {Phys. Rev. D}\ }\textbf {\bibinfo {volume}
  {49}},\ \bibinfo {pages} {2658} (\bibinfo {year} {1994})}\BibitemShut
  {NoStop}%
\bibitem [{\citenamefont {Veitch}\ and\ \citenamefont
  {Vecchio}(2010)}]{VeitchVecchio:2010}%
  \BibitemOpen
  \bibfield  {author} {\bibinfo {author} {\bibfnamefont {J.}~\bibnamefont
  {Veitch}}\ and\ \bibinfo {author} {\bibfnamefont {A.}~\bibnamefont
  {Vecchio}},\ }\href {\doibase 10.1103/PhysRevD.81.062003} {\bibfield
  {journal} {\bibinfo  {journal} {Phys. Rev. D}\ }\textbf {\bibinfo {volume}
  {81}},\ \bibinfo {pages} {062003} (\bibinfo {year} {2010})}\BibitemShut
  {NoStop}%
\bibitem [{\citenamefont {Littenberg}\ and\ \citenamefont
  {Cornish}(2009)}]{LittenbergCornish:2009}%
  \BibitemOpen
  \bibfield  {author} {\bibinfo {author} {\bibfnamefont {T.~B.}\ \bibnamefont
  {Littenberg}}\ and\ \bibinfo {author} {\bibfnamefont {N.~J.}\ \bibnamefont
  {Cornish}},\ }\href {\doibase 10.1103/PhysRevD.80.063007} {\bibfield
  {journal} {\bibinfo  {journal} {Phys. Rev. D}\ }\textbf {\bibinfo {volume}
  {80}},\ \bibinfo {pages} {063007} (\bibinfo {year} {2009})}\BibitemShut
  {NoStop}%
\bibitem [{\citenamefont {Poisson}\ and\ \citenamefont
  {Will}(1995)}]{PoissonWill}%
  \BibitemOpen
  \bibfield  {author} {\bibinfo {author} {\bibfnamefont {E.}~\bibnamefont
  {Poisson}}\ and\ \bibinfo {author} {\bibfnamefont {C.~M.}\ \bibnamefont
  {Will}},\ }\href {\doibase 10.1103/PhysRevD.52.848} {\bibfield  {journal}
  {\bibinfo  {journal} {Phys.Rev.}\ }\textbf {\bibinfo {volume} {D52}},\
  \bibinfo {pages} {848} (\bibinfo {year} {1995})},\ \Eprint
  {http://arxiv.org/abs/gr-qc/9502040} {arXiv:gr-qc/9502040} \BibitemShut
  {NoStop}%
\bibitem [{\citenamefont {Vallisneri}(2008)}]{Vallisneri:2007ev}%
  \BibitemOpen
  \bibfield  {author} {\bibinfo {author} {\bibfnamefont {M.}~\bibnamefont
  {Vallisneri}},\ }\href {\doibase 10.1103/PhysRevD.77.042001} {\bibfield
  {journal} {\bibinfo  {journal} {Phys.Rev.}\ }\textbf {\bibinfo {volume}
  {D77}},\ \bibinfo {pages} {042001} (\bibinfo {year} {2008})},\ \Eprint
  {http://arxiv.org/abs/gr-qc/0703086} {arXiv:gr-qc/0703086} \BibitemShut
  {NoStop}%
\bibitem [{\citenamefont {Cornish}\ and\ \citenamefont
  {Littenberg}(2007)}]{CornishLittenberg:2007}%
  \BibitemOpen
  \bibfield  {author} {\bibinfo {author} {\bibfnamefont {N.~J.}\ \bibnamefont
  {Cornish}}\ and\ \bibinfo {author} {\bibfnamefont {T.~B.}\ \bibnamefont
  {Littenberg}},\ }\href {\doibase 10.1103/PhysRevD.76.083006} {\bibfield
  {journal} {\bibinfo  {journal} {Phys. Rev. D}\ }\textbf {\bibinfo {volume}
  {76}},\ \bibinfo {pages} {083006} (\bibinfo {year} {2007})}\BibitemShut
  {NoStop}%
\bibitem [{\citenamefont {{R{\"o}ver}}(2010)}]{Roever:2010}%
  \BibitemOpen
  \bibfield  {author} {\bibinfo {author} {\bibfnamefont {C.}~\bibnamefont
  {{R{\"o}ver}}},\ }\href {\doibase 10.1088/1742-6596/228/1/012008} {\bibfield
  {journal} {\bibinfo  {journal} {Journal of Physics Conference Series}\
  }\textbf {\bibinfo {volume} {228}},\ \bibinfo {eid} {012008} (\bibinfo {year}
  {2010})},\ \Eprint {http://arxiv.org/abs/0911.5051} {arXiv:0911.5051}
  \BibitemShut {NoStop}%
\bibitem [{\citenamefont {Robinson}\ \emph {et~al.}(2008)\citenamefont
  {Robinson}, \citenamefont {Sathyaprakash},\ and\ \citenamefont
  {Sengupta}}]{Robinson:2008un}%
  \BibitemOpen
  \bibfield  {author} {\bibinfo {author} {\bibfnamefont {C.}~\bibnamefont
  {Robinson}}, \bibinfo {author} {\bibfnamefont {B.}~\bibnamefont
  {Sathyaprakash}}, \ and\ \bibinfo {author} {\bibfnamefont {A.~S.}\
  \bibnamefont {Sengupta}},\ }\href {\doibase 10.1103/PhysRevD.78.062002}
  {\bibfield  {journal} {\bibinfo  {journal} {Phys.Rev.}\ }\textbf {\bibinfo
  {volume} {D78}},\ \bibinfo {pages} {062002} (\bibinfo {year} {2008})},\
  \Eprint {http://arxiv.org/abs/0804.4816} {arXiv:0804.4816} \BibitemShut
  {NoStop}%
\bibitem [{\citenamefont {Schutz}(2011)}]{Schutz:2011tw}%
  \BibitemOpen
  \bibfield  {author} {\bibinfo {author} {\bibfnamefont {B.~F.}\ \bibnamefont
  {Schutz}},\ }\href {\doibase 10.1088/0264-9381/28/12/125023} {\bibfield
  {journal} {\bibinfo  {journal} {Class.Quant.Grav.}\ }\textbf {\bibinfo
  {volume} {28}},\ \bibinfo {pages} {125023} (\bibinfo {year} {2011})},\
  \Eprint {http://arxiv.org/abs/1102.5421} {arXiv:1102.5421} \BibitemShut
  {NoStop}%
\bibitem [{\citenamefont {Abadie}\ \emph {et~al.}(2012)\citenamefont {Abadie}
  \emph {et~al.}}]{Colaboration:2011np}%
  \BibitemOpen
  \bibfield  {author} {\bibinfo {author} {\bibfnamefont {J.}~\bibnamefont
  {Abadie}} \emph {et~al.} (\bibinfo {collaboration} {LIGO Collaboration, Virgo
  Collaboration}),\ }\href {\doibase 10.1103/PhysRevD.85.082002} {\bibfield
  {journal} {\bibinfo  {journal} {Phys. Rev.}\ }\textbf {\bibinfo {volume}
  {D85}},\ \bibinfo {pages} {082002} (\bibinfo {year} {2012})},\ \Eprint
  {http://arxiv.org/abs/1111.7314} {arXiv:1111.7314} \BibitemShut {NoStop}%
\bibitem [{\citenamefont {Robinson}\ \emph {et~al.}(2007)\citenamefont
  {Robinson}, \citenamefont {Sathyaprakash},\ and\ \citenamefont
  {Sengupta}}]{TrigScan}%
  \BibitemOpen
  \bibfield  {author} {\bibinfo {author} {\bibfnamefont {C.}~\bibnamefont
  {Robinson}}, \bibinfo {author} {\bibfnamefont {B.}~\bibnamefont
  {Sathyaprakash}}, \ and\ \bibinfo {author} {\bibfnamefont {A.}~\bibnamefont
  {Sengupta}},\ }\href {https://dcc.ligo.org/LIGO-G070460/public} {\emph
  {\bibinfo {title} {Taking parameter correlations into account in the binary
  inspiral pipeline}}},\ \bibinfo {type} {LIGO-DCC}\ \bibinfo {number}
  {G070460-00}\ (\bibinfo  {institution} {LIGO},\ \bibinfo {year}
  {2007})\BibitemShut {NoStop}%
\bibitem [{\citenamefont {Fairhurst}\ and\ \citenamefont
  {Brady}(2008)}]{Fairhurst:2007qj}%
  \BibitemOpen
  \bibfield  {author} {\bibinfo {author} {\bibfnamefont {S.}~\bibnamefont
  {Fairhurst}}\ and\ \bibinfo {author} {\bibfnamefont {P.}~\bibnamefont
  {Brady}},\ }\href {\doibase 10.1088/0264-9381/25/10/105002} {\bibfield
  {journal} {\bibinfo  {journal} {Class.Quant.Grav.}\ }\textbf {\bibinfo
  {volume} {25}},\ \bibinfo {pages} {105002} (\bibinfo {year} {2008})},\
  \Eprint {http://arxiv.org/abs/0707.2410} {arXiv:0707.2410} \BibitemShut
  {NoStop}%
\bibitem [{\citenamefont {Buonanno}\ \emph {et~al.}(2009)\citenamefont
  {Buonanno}, \citenamefont {Iyer}, \citenamefont {Ochsner}, \citenamefont
  {Pan},\ and\ \citenamefont {Sathyaprakash}}]{BIOPS}%
  \BibitemOpen
  \bibfield  {author} {\bibinfo {author} {\bibfnamefont {A.}~\bibnamefont
  {Buonanno}}, \bibinfo {author} {\bibfnamefont {B.~R.}\ \bibnamefont {Iyer}},
  \bibinfo {author} {\bibfnamefont {E.}~\bibnamefont {Ochsner}}, \bibinfo
  {author} {\bibfnamefont {Y.}~\bibnamefont {Pan}}, \ and\ \bibinfo {author}
  {\bibfnamefont {B.~S.}\ \bibnamefont {Sathyaprakash}},\ }\href {\doibase
  10.1103/PhysRevD.80.084043} {\bibfield  {journal} {\bibinfo  {journal} {Phys.
  Rev. D}\ }\textbf {\bibinfo {volume} {80}},\ \bibinfo {pages} {084043}
  (\bibinfo {year} {2009})}\BibitemShut {NoStop}%
\bibitem [{\citenamefont {{Nitz}}\ \emph {et~al.}(2013)\citenamefont {{Nitz}},
  \citenamefont {{Lundgren}}, \citenamefont {{Brown}}, \citenamefont
  {{Ochsner}}, \citenamefont {{Keppel}},\ and\ \citenamefont
  {{Harry}}}]{Nitz:2013}%
  \BibitemOpen
  \bibfield  {author} {\bibinfo {author} {\bibfnamefont {A.~H.}\ \bibnamefont
  {{Nitz}}}, \bibinfo {author} {\bibfnamefont {A.}~\bibnamefont {{Lundgren}}},
  \bibinfo {author} {\bibfnamefont {D.~A.}\ \bibnamefont {{Brown}}}, \bibinfo
  {author} {\bibfnamefont {E.}~\bibnamefont {{Ochsner}}}, \bibinfo {author}
  {\bibfnamefont {D.}~\bibnamefont {{Keppel}}}, \ and\ \bibinfo {author}
  {\bibfnamefont {I.~W.}\ \bibnamefont {{Harry}}},\ }\href@noop {} {\bibfield
  {journal} {\bibinfo  {journal} {ArXiv e-prints}\ } (\bibinfo {year}
  {2013})},\ \Eprint {http://arxiv.org/abs/1307.1757} {arXiv:1307.1757}
  \BibitemShut {NoStop}%
\bibitem [{LAL()}]{LALSuite}%
  \BibitemOpen
  \href@noop {} {}\bibinfo {howpublished}
  {\url{http://www.lsc-group.phys.uwm.edu/daswg/projects/lalsuite.html}\,}\BibitemShut
  {NoStop}%
\bibitem [{\citenamefont {Aasi}\ \emph {et~al.}(2013)\citenamefont {Aasi} \emph
  {et~al.}}]{Aasi:2013wya}%
  \BibitemOpen
  \bibfield  {author} {\bibinfo {author} {\bibfnamefont {J.}~\bibnamefont
  {Aasi}} \emph {et~al.} (\bibinfo {collaboration} {LIGO Scientific
  Collaboration, Virgo Collaboration}),\ }\href@noop {} {\  (\bibinfo {year}
  {2013})},\ \Eprint {http://arxiv.org/abs/1304.0670} {arXiv:1304.0670}
  \BibitemShut {NoStop}%
\bibitem [{\citenamefont {Droz}\ \emph {et~al.}(1999)\citenamefont {Droz},
  \citenamefont {Knapp}, \citenamefont {Poisson},\ and\ \citenamefont
  {Owen}}]{DrozSPA:1999}%
  \BibitemOpen
  \bibfield  {author} {\bibinfo {author} {\bibfnamefont {S.}~\bibnamefont
  {Droz}}, \bibinfo {author} {\bibfnamefont {D.~J.}\ \bibnamefont {Knapp}},
  \bibinfo {author} {\bibfnamefont {E.}~\bibnamefont {Poisson}}, \ and\
  \bibinfo {author} {\bibfnamefont {B.~J.}\ \bibnamefont {Owen}},\ }\href
  {\doibase 10.1103/PhysRevD.59.124016} {\bibfield  {journal} {\bibinfo
  {journal} {Phys. Rev. D}\ }\textbf {\bibinfo {volume} {59}},\ \bibinfo
  {pages} {124016} (\bibinfo {year} {1999})}\BibitemShut {NoStop}%
\bibitem [{Ran()}]{RandomForest}%
  \BibitemOpen
  \href@noop {} {}\bibinfo {howpublished}
  {\url{http://www.stat.berkeley.edu/~breiman/RandomForests/cc_home.htm}\,}\BibitemShut
  {NoStop}%
\bibitem [{\citenamefont {Pedregosa}\ \emph {et~al.}(2011)\citenamefont
  {Pedregosa} \emph {et~al.}}]{scikit-learn}%
  \BibitemOpen
  \bibfield  {author} {\bibinfo {author} {\bibfnamefont {F.}~\bibnamefont
  {Pedregosa}} \emph {et~al.},\ }\href
  {http://dl.acm.org/citation.cfm?id=1953048.2078195} {\bibfield  {journal}
  {\bibinfo  {journal} {Journal of Machine Learning Research}\ }\textbf
  {\bibinfo {volume} {12}},\ \bibinfo {pages} {2825} (\bibinfo {year}
  {2011})}\BibitemShut {NoStop}%
\bibitem [{\citenamefont {Dominik}\ \emph {et~al.}(2012)\citenamefont {Dominik}
  \emph {et~al.}}]{DominikEtAl:2012}%
  \BibitemOpen
  \bibfield  {author} {\bibinfo {author} {\bibfnamefont {M.}~\bibnamefont
  {Dominik}} \emph {et~al.},\ }\href {\doibase 10.1088/0004-637X/759/1/52}
  {\bibfield  {journal} {\bibinfo  {journal} {The Astrophysical Journal}\
  }\textbf {\bibinfo {volume} {759}},\ \bibinfo {pages} {52} (\bibinfo {year}
  {2012})}\BibitemShut {NoStop}%
\bibitem [{\citenamefont {Cannon}(2008)}]{Cannon:2008zz}%
  \BibitemOpen
  \bibfield  {author} {\bibinfo {author} {\bibfnamefont {K.~C.}\ \bibnamefont
  {Cannon}},\ }\href {\doibase 10.1088/0264-9381/25/10/105024} {\bibfield
  {journal} {\bibinfo  {journal} {Class. Quant. Grav.}\ }\textbf {\bibinfo
  {volume} {25}},\ \bibinfo {pages} {105024} (\bibinfo {year}
  {2008})}\BibitemShut {NoStop}%
\bibitem [{\citenamefont {Blackburn}\ \emph {et~al.}(2008)\citenamefont
  {Blackburn}, \citenamefont {Cadonati}, \citenamefont {Caride}, \citenamefont
  {Caudill}, \citenamefont {Chatterji} \emph {et~al.}}]{Blackburn:2008ah}%
  \BibitemOpen
  \bibfield  {author} {\bibinfo {author} {\bibfnamefont {L.}~\bibnamefont
  {Blackburn}}, \bibinfo {author} {\bibfnamefont {L.}~\bibnamefont {Cadonati}},
  \bibinfo {author} {\bibfnamefont {S.}~\bibnamefont {Caride}}, \bibinfo
  {author} {\bibfnamefont {S.}~\bibnamefont {Caudill}}, \bibinfo {author}
  {\bibfnamefont {S.}~\bibnamefont {Chatterji}},  \emph {et~al.},\ }\href
  {\doibase 10.1088/0264-9381/25/18/184004} {\bibfield  {journal} {\bibinfo
  {journal} {Class.Quant.Grav.}\ }\textbf {\bibinfo {volume} {25}},\ \bibinfo
  {pages} {184004} (\bibinfo {year} {2008})},\ \Eprint
  {http://arxiv.org/abs/0804.0800} {arXiv:0804.0800} \BibitemShut {NoStop}%
\bibitem [{\citenamefont {Allen}(2005)}]{Allen:2004gu}%
  \BibitemOpen
  \bibfield  {author} {\bibinfo {author} {\bibfnamefont {B.}~\bibnamefont
  {Allen}},\ }\href {\doibase 10.1103/PhysRevD.71.062001} {\bibfield  {journal}
  {\bibinfo  {journal} {Phys.Rev.}\ }\textbf {\bibinfo {volume} {D71}},\
  \bibinfo {pages} {062001} (\bibinfo {year} {2005})},\ \Eprint
  {http://arxiv.org/abs/gr-qc/0405045} {arXiv:gr-qc/0405045} \BibitemShut
  {NoStop}%
\bibitem [{\citenamefont {Hanna}(2008)}]{ChadThesis}%
  \BibitemOpen
  \bibfield  {author} {\bibinfo {author} {\bibfnamefont {C.}~\bibnamefont
  {Hanna}},\ }\emph {\bibinfo {title} {Searching for Gravitational Waves from
  Binary Systems in Non-Stationary Data}},\ \href
  {http://etd.lsu.edu/docs/available/etd-03272008-092832/} {Ph.D. thesis},\
  \bibinfo  {school} {Louisiana State University} (\bibinfo {year}
  {2008})\BibitemShut {NoStop}%
\bibitem [{\citenamefont {Allen}\ \emph {et~al.}(2003)\citenamefont {Allen},
  \citenamefont {Creighton}, \citenamefont {Flanagan},\ and\ \citenamefont
  {Romano}}]{Allen:2002jw}%
  \BibitemOpen
  \bibfield  {author} {\bibinfo {author} {\bibfnamefont {B.}~\bibnamefont
  {Allen}}, \bibinfo {author} {\bibfnamefont {J.~D.}\ \bibnamefont
  {Creighton}}, \bibinfo {author} {\bibfnamefont {E.~E.}\ \bibnamefont
  {Flanagan}}, \ and\ \bibinfo {author} {\bibfnamefont {J.~D.}\ \bibnamefont
  {Romano}},\ }\href {\doibase 10.1103/PhysRevD.67.122002} {\bibfield
  {journal} {\bibinfo  {journal} {Phys. Rev.}\ }\textbf {\bibinfo {volume}
  {D67}},\ \bibinfo {pages} {122002} (\bibinfo {year} {2003})},\ \Eprint
  {http://arxiv.org/abs/gr-qc/0205015} {arXiv:gr-qc/0205015} \BibitemShut
  {NoStop}%
\bibitem [{\citenamefont {Littenberg}\ and\ \citenamefont
  {Cornish}(2010)}]{Littenberg2010}%
  \BibitemOpen
  \bibfield  {author} {\bibinfo {author} {\bibfnamefont {T.~B.}\ \bibnamefont
  {Littenberg}}\ and\ \bibinfo {author} {\bibfnamefont {N.~J.}\ \bibnamefont
  {Cornish}},\ }\href {\doibase 10.1103/PhysRevD.82.103007} {\bibfield
  {journal} {\bibinfo  {journal} {Phys. Rev.}\ }\textbf {\bibinfo {volume}
  {D82}},\ \bibinfo {pages} {103007} (\bibinfo {year} {2010})},\ \Eprint
  {http://arxiv.org/abs/1008.1577} {arXiv:1008.1577} \BibitemShut {NoStop}%
\bibitem [{\citenamefont {Searle}(2008)}]{Searle:2008}%
  \BibitemOpen
  \bibfield  {author} {\bibinfo {author} {\bibfnamefont {A.~C.}\ \bibnamefont
  {Searle}},\ }\href@noop {} {\  (\bibinfo {year} {2008})},\ \Eprint
  {http://arxiv.org/abs/0804.1161} {arXiv:0804.1161} \BibitemShut {NoStop}%
\end{thebibliography}%

\end{document}